\documentclass[]{spie}  
\usepackage[]{graphicx,bbold}
\usepackage[]{subeqn,url}
\title{Slepian Functions and Their Use\\ in Signal Estimation and
Spectral Analysis}
\author{Frederik J.~Simons
\skiplinehalf
Department of Geosciences, Princeton University, Guyot Hall,
Princeton, NJ, USA
}
\newcommand{\WS}{^{\mathrm{WS}}}
\newcommand{\SP}{^{\mathrm{SP}}}
\newcommand{\MT}{^{\mathrm{MT}}}

\newcommand{\pinv}{{\mathrm{pinv}}}
\newcommand{\bxi}{\mbox{\boldmath$\xi$}}
\newcommand{\intinf}{\int_{-\infty}^{\infty}}

\newcommand{\intW}{\int_{-W}^{W}}
\newcommand{\sumshLp}{\suml_{l'=0}^{L}\suml_{m'=-l'}^{l'}}
\newcommand{\intR}{\int_\mathcal{R}}
\newcommand{\D}{^{\mathrm{1D}}}
\newcommand{\DD}{^{\mathrm{2D}}}
\newcommand{\DDD}{^{\mathrm{3D}}}
\newcommand{\Dxxp}{D(\bx,\bx')}
\newcommand{\Dkkp}{D(\bk,\bk')}
\newcommand{\Dttp}{D(t,t')}
\newcommand{\Drhrhp}{D(\rhat,\rhat')}

\newcommand{\Dxixip}{D(\bxi,\bxi')}
\newcommand{\Dxixi}{D(\bxi,\bxi)}
\newcommand{\fnorm}{(2\pi)^{-2}}
\newcommand{\jnorm}{(2\pi)^{-1}}

\newcommand{\intK}{\int_\mathcal{K}}

\newcommand{\bx}{\mathbf{x}}

\newcommand{\bk}{\mathbf{k}}
\newcommand{\dbk}{\,d\bk}

\newcommand{\dbx}{\,d\bx}

\newcommand{\domg}{\,d\Omega} 
\newcommand{\fracd}[2]{\frac{\displaystyle{#1}}{\displaystyle{#2}}} 
\newcommand{\intr}{\int_R} 
\newcommand{\intbr}{\int_{\bar{R}}} 
 
\newcommand{\into}{\int_\Omega}

\newcommand{\also}{\quad\mbox{and}\quad} 
\newcommand{\with}{\quad\mbox{with}\quad} 
 
\newcommand{\for}{\quad\mbox{for}\quad} 
\newcommand{\dab}{\delta_{\alpha\beta}}

\newcommand{\dllp}{\delta_{ll'}} 
\newcommand{\dmmp}{\delta_{mm'}} 
\newcommand{\Ylm}{Y_{lm}} 
\newcommand{\Ylmrh}{Y_{lm}(\rhat)} 
\newcommand{\Ylmrhp}{Y_{lm}(\rhat')}

\newcommand{\Ylmp}{Y_{l'm'}} 
\newcommand{\Xlm}{X_{lm}}

\newcommand{\Xlmth}{X_{lm}(\theta)} 
\newcommand{\Xlmthp}{X_{lm}(\theta')}

\newcommand{\Dlmlmp}{D_{lm,l'm'}} 
\newcommand{\bDlmlmp}{\bar{D}_{lm,l'm'}}

\newcommand{\diag}{\textrm{diag}}

\newcommand{\grhp}{g(\rhat')} 
\newcommand{\garh}{g_\alpha(\rhat)}

\newcommand{\glma}{g_{lm \hsp\alpha}} 
 
\newcommand{\galm}{\glma} 
 
\newcommand{\glmb}{g_{lm\hsp\beta}} 
\newcommand{\glmpb}{g_{l'm'\hsp\beta}} 
\newcommand{\gblmp}{\glmpb} 
 
\newcommand{\Lpot}{(L+1)^2} 
\newcommand{\sumapot}{\sum_{\alpha=1}^{\Lpot}} 
 
\newcommand{\sumakR}{\sum_{\alpha>N\DDD}^{\Lpot}} 
\newcommand{\sumaN}{\sum_{\alpha=1}^{N\DDD}}

\newcommand{\suml}{\sum\limits}

\newcommand{\sumsh}{\suml_{l=0}^{\infty}\suml_{m=-l}^{l}} 
\newcommand{\sumshL}{\suml_{l=0}^{L}\suml_{m=-l}^{l}}
\newcommand{\tlofp}{\left(\frac{2l+1}{4\pi}\right)} 
\newcommand{\Plm}{P_{lm}}

\newcommand{\blambda}{\mbox{\boldmath$\Lambda$}} 
\newcommand{\bsigma}{\mbox{\boldmath$\Sigma$}} 
 
\newcommand{\rhat}{\mbf{\hat{r}}} 
\newcommand{\shat}{\hat{s}} 
\newcommand{\Shat}{\hat{S}} 
 
\newcommand{\be}{\begin{equation}} 
\newcommand{\ee}{\end{equation}} 
\newcommand{\ber}{\begin{eqnarray}} 
\newcommand{\eer}{\end{eqnarray}} 
\newcommand{\barray}{\begin{array}} 
\newcommand{\earray}{\end{array}} 
\newcommand{\mbf}{\mathbf}

\newcommand{\nnr}{\nonumber}
\newcommand{\sst}{\scriptstyle} 

\newcommand{\rar}{\rightarrow} 
\newcommand{\ssec}{\subsection} 
\newcommand{\sssec}{\subsubsection} 
 
\newcommand{\hsp}{\hspace*{0.1em}}

\newcommand{\hspm}{\hspace*{-0.1em}}

\newcommand{\beff}{\mbox{\boldmath$\mathsf{f}$}}
\newcommand{\begg}{\mbox{\boldmath$\mathsf{G}$}}
\newcommand{\beggbar}{\mbox{\boldmath$\bar{\mathsf{G}}$}}
\newcommand{\beggubar}{\mbox{\boldmath{\b{$\mathsf{G}$}}}}
\newcommand{\blambdaubar}{\mbox{\boldmath{\b{$\Lambda$}}}} 
\newcommand{\blambdabar}{\mbox{\boldmath$\bar{\Lambda}$}} 
\newcommand{\Cbubar}{\mbox{\boldmath{\b{$\mathrm{B}$}}}}
\newcommand{\Wbubar}{\mbox{\boldmath{\b{$\mathrm{V}$}}}}
\newcommand{\Ibubar}{\mbox{\boldmath{\b{$\mathrm{I}$}}}}
\newcommand{\bY}{\mbox{\boldmath$\mathsf{Y}$}}
\newcommand{\bG}{\mbox{\boldmath$\mathsf{G}$}}
\newcommand{\br}{\rhat}
\newcommand{\Tit}{^{\it{\sst{T}}}}
\newcommand{\Trm}{^{\mathrm{\sst{T}}}}
\newcommand{\T}{^{\sf{\sst{T}}}}
\newcommand{\fb}{\mathbf{f}}
\newcommand{\fbh}{\mathbf{\hat{f}}}
\newcommand{\tb}{\mathbf{t}}
\newcommand{\tbh}{\mathbf{\hat{t}}}
\newcommand{\Gb}{\mathbf{G}}
\newcommand{\Ub}{\mathbf{U}}
\newcommand{\Db}{\mathbf{D}}
\newcommand{\Ib}{\mathbf{I}}
\newcommand{\Ibbar}{\bar{\mathbf{I}}}
\newcommand{\Hb}{\mathbf{H}}
\newcommand{\Vb}{\mathbf{V}}
\newcommand{\Bb}{\mathbf{B}}

\newcommand{\Cbbar}{\mathbf{\bar{B}}}

\newcommand{\Wbbar}{\mathbf{\bar{V}}}
\newcommand{\sumshortL}{\suml_{lm}^{L}}
\newcommand{\sumshortLp}{\suml_{l'm'}^{L}}
\newcommand{\bd}{\mbox{\boldmath$\mathsf{d}$}}

\newcommand{\bP}{\mbox{\boldmath$\mathsf{P}$}}

\begin{document} 
\maketitle 

\begin{abstract}
\end{abstract}

It is a well-known fact that mathematical functions
that are timelimited (or spacelimited) cannot be simultaneously
bandlimited (in frequency). Yet the finite precision of
measurement and computation unavoidably bandlimits our observation and
modeling scientific data, and we often only have access to, or are
only interested in, a study area that is temporally or spatially
bounded. In the geosciences we may be interested in spectrally
modeling a time series defined only on a certain interval, or we may
want to characterize a specific geographical area observed using an
effectively bandlimited measurement device. It is clear that analyzing
and representing scientific data of this kind will be facilitated if a
basis of functions can be found that are ``spatiospectrally''
concentrated, i.e. ``localized'' in both domains at the same
time. Here, we give a  theoretical overview of one particular
approach to this ``concentration'' problem, as originally proposed for
time series by Slepian and coworkers, in the 1960s. We show how this
framework leads to practical algorithms and statistically performant
methods for the analysis of signals and their power spectra in one and
two dimensions, and on the surface of a sphere.


\section{INTRODUCTION}

It is well appreciated that functions cannot have
finite support in the temporal (or spatial) and spectral domain at the
same time\cite{Slepian83}. Finding and representing signals  that
are optimally concentrated in both is a fundamental problem in
information theory  which was solved in the early 1960s by Slepian,  
Landau and Pollak\cite{Slepian+61,Landau+61,Landau+62}. The
extensions and generalizations of this 
problem\cite{Daubechies88,Daubechies+88,Cohen89,Daubechies90} have
strong connections with the  burgeoning field of wavelet
analysis.  In this contribution, however, we shall \textit{not} talk
about wavelets, the scaled translates of a ``mother'' with vanishing
moments, the tool for multi-resolution
analysis\cite{Daubechies92,Flandrin98,Mallat98}. Rather, we devote our
attention entirely to what we shall collectively refer to as ``Slepian
functions'', in multiple Cartesian dimensions and on the sphere.   

These we understand to be orthogonal families
of functions that are all defined on a common, e.g. geographical,
domain, where they are either optimally concentrated or within which they
are exactly limited, and which at the same time are exactly confined
within a certain bandwidth, or maximally concentrated therein. The
measure of concentration is invariably a quadratic energy ratio, which,
though only one choice out of many\cite{Donoho+89,Freeden+97,Riedel+95} is
perfectly suited to the nature of the  problems we are attempting to
address. These are, for example: How do we make estimates of signals
that are noisily and incompletely observed? How do we analyze the
properties of  such signals efficiently, and how can we represent them
economically? How do we estimate the power spectrum of noisy and
incomplete data? What are the particular constraints imposed by
dealing with potential-field signals (gravity, magnetism, etc) and how
is the altitude of the observation point, e.g. from a satellite in
orbit, taken into account? What are the statistical properties of the
resulting signal and power spectral estimates?    

These and other questions have been studied extensively in one
dimension, that is, for time series, but remarkably little work had been
done in the Cartesian plane or on the surface of the
sphere. For the geosciences, the latter two domains of application are
nevertheless vital for the obvious reasons that they deal with
information (measurement and modeling) that is geographically
distributed on (a portion of) a planetary surface. In our own recent
series of
papers\cite{Wieczorek+2005,Simons+2006a,Simons+2006b,Simons+2007,Wieczorek+2007,Dahlen+2008,Simons+2009}
we have dealt extensively with Slepian's problem in spherical
geometry; asymptotic reductions to the
plane\cite{Simons+2006a,Simons+2010} then generalize Slepian's early 
treatment of the multidimensional Cartesian case\cite{Slepian64}. 

In this chapter we provide a framework for the analysis and
representation of geoscientific data by means of Slepian
functions defined for time series, on two-dimensional Cartesian, and
spherical domains. We emphasize the common ground underlying the
construction of all Slepian functions, discuss practical algorithms,
and review the major findings of our own recent work on
signal\cite{Wieczorek+2005,Simons+2006b} and power spectral
estimation theory on the sphere\cite{Wieczorek+2007,Dahlen+2008}.    

\section{THEORY OF SLEPIAN FUNCTIONS}
 
In this section we review the theory of Slepian functions in one
dimension, in the Cartesian plane, and on the surface of the unit
sphere. The one-dimensional theory is quite well known and perhaps
most accessibly presented in the textbook by Percival and
Walden\cite{Percival+93}. It is briefly reformulated here for
consistency and to establish some notation. The two-dimensional planar
case formed the subject of a lesser-known of Slepian's
papers\cite{Slepian64} and is reviewed here also. We are not
discussing alternatives by which two-dimensional Slepian functions are
constructed by forming the outer product of pairs of one-dimensional
functions. While this approach has produced some useful
results\cite{Hanssen97,Simons+2000}, it does not solve the
concentration problem \textit{sensu stricto}. The spherical
case was treated in most detail, and for the first time, by ourselves
elsewhere\cite{Wieczorek+2005,Simons+2006a,Simons+2006b}, though two
very important early studies by Slepian, Gr\"unbaum, and others,
laid much of the foundation for the analytical treatment of the
spherical concentration problem for cases with special symmetries
\cite{Gilbert+77,Grunbaum81a}. Finally, we recast the theory in
the context of reproducing-kernel Hilbert spaces, through which the
reader may appreciate some of the connections with radial basis
functions, splines, and  wavelet analysis, which are commonly
formulated in such a framework\cite{Freeden+98c}.  

\ssec{Spatiospectral concentration for time series}

\ssec*{General theory in one dimension}

We use~$t$ to denote time or one-dimensional space and~$\omega$ for
angular frequency, and adopt a normalization convention\cite{Mallat98}
in which a real-valued time-domain signal~$f(t)$ and its Fourier
transform~$F(\omega)$ are related by 
\be
\label{slepian0}
f(t)=\jnorm\intinf 
F(\omega)e^{i\omega t}\,d\omega,\qquad 
F(\omega)=\intinf 
f(t)e^{-i\omega t}\,dt.
\ee
The problem of finding the strictly bandlimited signal 
\be
\label{gband}
g(t)=\jnorm\intW 
G(\omega)e^{i\omega t}\,d\omega
,
\ee
that is maximally (though by virtue of the Paley-Wiener
theorem\cite{Daubechies92,Mallat98} never completely) concentrated
into a time interval $|t|\le T$ was first considered by Slepian, Landau and
Pollak\cite{Slepian+61,Landau+61}. The optimally concentrated signal
is taken to be the one with the least energy outside of the interval:
\be
\lambda=\fracd{\int_{-T}^{T}g^2(t)\,dt}
{\intinf g^2(t)\,dt}=\mbox{maximum}.
\label{slepian1}
\ee
Bandlimited functions~$g(t)$ satisfying the variational
problem~(\ref{slepian1}) have spectra~$G(\omega)$ that satisfy the
frequency-domain convolutional integral eigenvalue equation
\begin{subequations}
\label{slepian2}
\be
\intW D(\omega,\omega')\,G(\omega')\,d\omega'
=\lambda\hspace{0.05em}G(\omega),\quad |\omega|\le W,
\ee
\be
D(\omega,\omega')=\frac{\sin T(\omega-\omega')}
{\pi (\omega-\omega')}.
\ee
\end{subequations}
The corresponding time- or spatial-domain formulation is
\begin{subequations}
\label{dude1}
\be
\label{eig1}
\int_{-T}^{T}\Dttp\,g(t')\,dt'=\lambda\hspace{0.05em}g(t),\quad
t\in\mathbb{R}
,
\ee
\be
\label{slepian4}
\Dttp=\frac{\sin W(t-t')}{\pi (t-t')}
.
\ee
\end{subequations}
The ``prolate spheroidal eigentapers'' $g_1(t),g_2(t),\ldots$ that
solve eq.~(\ref{dude1}) form a
doubly orthogonal set. When they are chosen to be orthonormal over 
infinite time $|t|\le\infty$ they are also orthogonal over
the finite interval $|t|\le T$: 
\be
\intinf g_{\alpha}g_{\beta}\,dt=\delta_{\alpha\beta}
,\qquad\int_{-T}^{T}g_{\alpha}g_{\beta}\,dt=
\lambda_{\alpha}\hspace{0.05em}\delta_{\alpha\beta}.
\label{slepian8}
\ee
A change of variables and a scaling of the eigenfunctions transforms
eq.~(\ref{slepian2}) into the dimensionless eigenproblem
\begin{subequations}
\label{slepian5}
\be
\int_{-1}^{1}D(x,x')\,\psi(x')\,dx'=\lambda\hspace{0.05em}\psi(x),
\ee
\be
D(x,x')=\fracd{\sin TW(x-x')}{\pi (x-x')}.
\ee
\end{subequations}
Eq.~(\ref{slepian5}) shows that the eigenvalues 
$\lambda_1>\lambda_2>\ldots$ and suitably scaled eigenfunctions
$\psi_1(x),\psi_2(x),\ldots$ depend only upon the time-bandwidth
product~$TW$. The sum of the concentration eigenvalues~$\lambda$ 
relates to this product by  
\be
N\D=\sum_{\alpha=1}^{\infty}\lambda_{\alpha}
=\int_{-1}^{1}D(x,x)\,dx
=\frac{(2T)(2W)}{2\pi}= 
\frac{2TW}{\pi}. 
\label{slepian6}
\ee
The shape of the eigenvalue spectrum of eq.~(\ref{slepian5}) has a
characteristic step shape, showing significant ($\lambda\approx 1$)
and insignificant ($\lambda\approx 0$) eigenvalues separated by a
narrow transition band\cite{Landau65,Slepian+65}. Thus, this
``Shannon number'' is a good estimate of the number of
significant eigenvalues, or, roughly speaking, $N\D$ is the number of signals
$f(t)$ that can be simultaneously well concentrated into a finite time
interval $|t|\le T$ and a finite frequency interval $|\omega|\le
W$. In other words\cite{Landau+62}, $N\D$ is the approximate dimension
of the space of signals that is ``essentially'' timelimited to~$T$ and
bandlimited to~$W$, and using the orthogonal set
$g_1,g_2,\ldots,g_{N\D}$ as its basis is parsimonious. 

\sssec*{Sturm-Liouville character and tridiagonal matrix formulation}

The integral operator acting upon~$\psi$ in eq.~(\ref{slepian5})
commutes with a differential operator that arises in expressing the
three-dimensional scalar wave equation in prolate spheroidal
coordinates\cite{Slepian+61,Slepian83}, which makes it possible to
find the scaled eigenfunctions~$\psi$ by solving the
Sturm-Liouville equation  
\be
\label{slepian7.5}
\frac{d}{dx}\left[(1-x^2)\frac{d\psi}{dx}\right]+
\left[\chi-\frac{(N\D)^2\pi^2}{4}x^2\right]  
\hsp\psi=0, 
\ee
where~$\chi\ne\lambda$ is the associated eigenvalue. The
eigenfunctions~$\psi(x)$ of eq.~(\ref{slepian7.5}) can be found at
discrete values of~$x$ by diagonalization of a simple symmetric
tridiagonal matrix\cite{Slepian78,Grunbaum81a,Percival+93} with
elements 
\ber
([N-1-2x]/2)^2\cos(2\pi W)\for x=0,\cdots,N-1,\nnr\\
x(N-x)/2\for x=1,\ldots,N-1
.
\eer
The matching eigenvalues~$\lambda$ can then be obtained directly from
eq.~(\ref{slepian5}). The discovery of the Sturm-Liouville formulation
of the concentration problem posed in eq.~(\ref{slepian1}) proved to
be a major impetus for the widespread adoption and practical
applications of the ``Slepian'' basis in signal identification,
spectral analysis and numerical analysis. Compared to the sequence of
eigenvalues~$\lambda$, the spectrum of the eigenvalues~$\chi$ is
extremely regular and thus the solution of eq.~(\ref{slepian7.5}) is
without any problem amenable to finite-precision numerical
computation\cite{Percival+93}.

\ssec{Spatiospectral concentration in the Cartesian plane}

\ssec*{General theory in two dimensions}

A square-integrable function~$f(\bx)$ defined in
the plane has the two-dimensional Fourier representation 
\be
\label{Bfourier}
f(\bx)=\fnorm\intinf
F(\bk)e^{i\bk\cdot\bx}\dbk,\qquad
F(\bk)=\intinf
f(\bx)e^{-i\bk\cdot\bx}\dbx,
\ee
We use~$g(\bx)$ to denote a function that is bandlimited
to~$\mathcal{K}$, an arbitrary subregion of spectral space, 
\be
\label{Bgdefn}
g(\bx)=\fnorm\!\!
\intK G(\bk)e^{i\mbf{k}\cdot\bx}\dbk
.
\ee
Following Slepian\cite{Slepian64}, we seek to concentrate the power
of~$g(\bx)$ into a finite spatial region~$\mathcal{R}\in\mathbb{R}^2$,
of area~$A$: 
\be
\label{Bnormratio}
\lambda=\fracd{\intR g^2(\bx)\dbx}
{\intinf g^2(\bx)\dbx}=\mbox{maximum}.
\ee
Bandlimited functions~$g(\bx)$ that maximize the Rayleigh
quotient~(\ref{Bnormratio}) solve the Fredholm integral equation
\begin{subequations}
\label{Beigen12}
\be
\label{Beigen1}
\intK \Dkkp\,G(\bk')\dbk'
=\lambda\hsp G(\bk),\qquad\bk\in\mathcal{K}
,
\ee
\be
\label{Beigen2}
\Dkkp=\fnorm
\intR e^{i(\bk'-\bk)\cdot\bx}\dbx
.
\ee
\end{subequations}
The corresponding problem in the spatial domain is  
\begin{subequations}
\label{Beigen34}
\be
\label{eig2}
\intR\! \Dxxp\,g(\bx')\dbx'
=\lambda\hsp g(\bx),\qquad\bx\in\mathbb{R}^2,
\ee
\be
\label{Beigen4}
\Dxxp=\fnorm
\intK e^{i\bk\cdot(\bx-\bx')}\dbk.
\ee
\end{subequations}
The bandlimited spatial-domain eigenfunctions
$g_1(\bx),g_2(\bx),\ldots$ and eigenvalues $\lambda_1\ge\lambda_2\ge\ldots$
that solve eq.~(\ref{Beigen34}) may be chosen to be orthonormal over
the whole plane $\|\bx\|\le\infty$ in which case they are also
orthogonal over~$\mathcal{R}$: 
\be
\intinf
g_{\alpha}g_{\beta}\dbx=\delta_{\alpha\beta},\qquad
\label{Borthog1}
\intR g_{\alpha}g_{\beta}\dbx=\lambda_{\alpha}\delta_{\alpha\beta}.
\label{Borthog2}
\ee
Concentration to the disk-shaped spectral band
$\mathcal{K}=\{\bk:\|\bk\|\le K\}$ allows us to rewrite
eq.~(\ref{Beigen34}) after a change of variables and a 
scaling of the eigenfunctions as 
\begin{subequations}
\label{Dscaled}
\be
\int_{\mathcal{R}_{*}}\!\!\Dxixip
\,\psi(\bxi')\,d\bxi'=\lambda\hsp\psi(\bxi),
\ee
\be
\Dxixip=\frac{K\sqrt{A/4\pi}}{2\pi}
\fracd{J_1(K\sqrt{A/4\pi}\,\,\|\bxi-\bxi'\|)} 
{\|\bxi-\bxi'\|},
\ee
\end{subequations}
where the region~$\mathcal{R}_*$ is scaled to area~$4\pi$ and
$J_1$ is the first-order Bessel function of the first
kind. Eq.~(\ref{Dscaled}) shows that, also in the two-dimensional
case, the eigenvalues $\lambda_1,\lambda_2,\ldots$ and the scaled
eigenfunctions $\psi_1(\bxi),\psi_2(\bxi),\ldots$ depend only on the
combination of the circular bandwidth~$K$ and the spatial
concentration area~$A$, where the quantity $K^2A/(4\pi)$ now plays the
role of the time-bandwidth product~$TW$ in the one-dimensional
case. The sum of the concentration eigenvalues~$\lambda$ defines the 
two-dimensional Shannon number~$N\DD$ as 
\be
\label{Bshannon}
N\DD=\sum_{\alpha=1}^{\infty}\lambda_{\alpha}
=\int_{R_{*}}\!
\Dxixi\,d\bxi
=\fracd{(\pi K^2)(A)}{(2\pi)^2}
= K^2\,\fracd{A}{4\pi}
.
\ee
Just as~$N\D$ in eq.~(\ref{slepian6}), $N\DD$ is the product of
the spectral and spatial areas of concentration multiplied by the
``Nyquist density''\cite{Daubechies88,Daubechies92}. And, similarly, it
is the effective dimension of the space of ``essentially'' space- and
bandlimited functions in which the set of two-dimensional functions
$g_1,g_2,\dots,g_{N\DD}$ may act as a sparse orthogonal basis.   

\begin{figure}[htb]\center
\rotatebox{-90}{\includegraphics[height=\columnwidth]{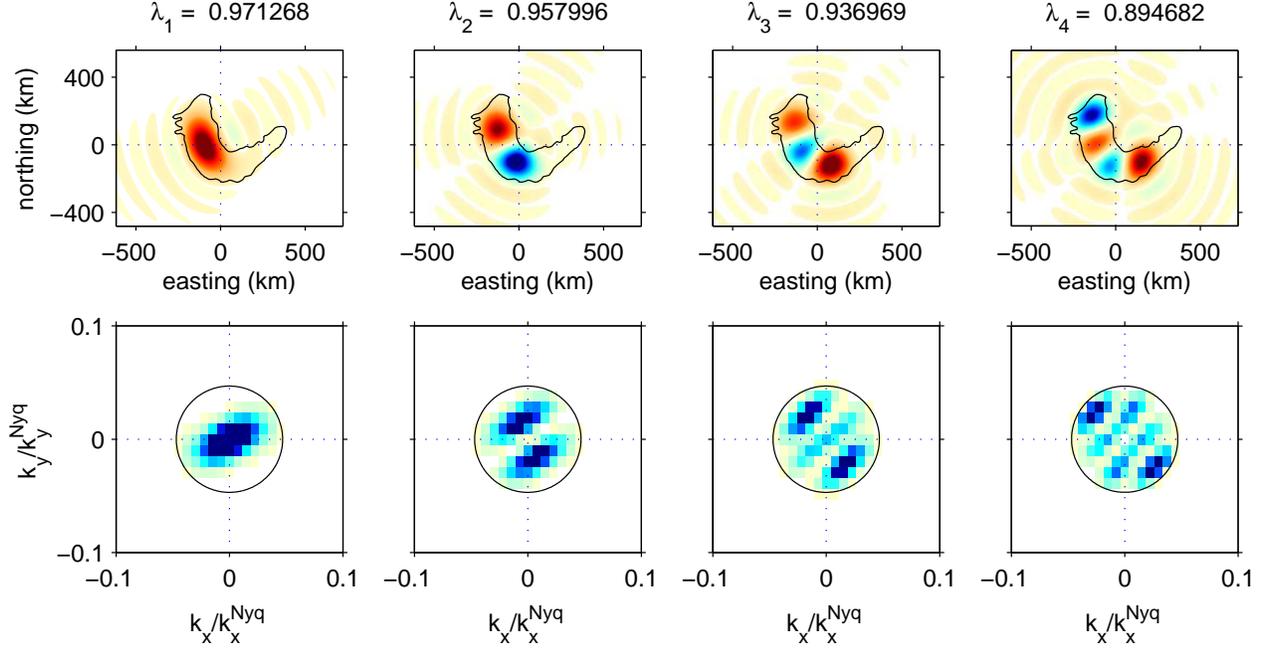}}
\caption{\label{swregions2d_2}Bandlimited eigenfunctions
$g_1,g_2,\ldots,g_{4}$ that are optimally concentrated within the
Columbia Plateau, a physiographic region in the United States centered 
on 116.02$^\circ$W 43.56$^\circ$N (near Boise City, Idaho) of area
$A\approx145\times 10^3$~km$^2$. The concentration factors
$\lambda_1,\lambda_2,\ldots,\lambda_{4}$ are indicated; the Shannon
number $N\DD=10$. The top row shows a rendition of the eigenfunctions
in space on a grid with 5~km resolution in both directions, with the
convention that positive values are blue and negative values red,
though the sign of the functions is arbitrary. The
spatial concentration region is outlined in black. The bottom row
shows the squared Fourier coefficients $|G_\alpha(\mbf{k})|^2$ as
calculated from the functions $g_\alpha(\bx)$ shown, on a wavenumber
scale that is expressed as a fraction of the Nyquist wavenumber. The
spectral limitation region is shown by the black circle at wavenumber
$K=0.0295$~rad/km. All areas for which the absolute value of the
functions plotted is less than one hundredth of the maximum value
attained over the domain are left white. The calculations were
performed by the Nystr\"om method using Gauss-Legendre integration of
eq.~(\ref{Dscaled}) in the two-dimensional spatial domain\cite{Simons+2010}.}
\end{figure}

An example of Slepian functions on a geographical domain
in the Cartesian plane can be found in Figure~\ref{swregions2d_2}.

\sssec*{Sturm-Liouville character and tridiagonal matrix formulation}

If in addition to the circular spectral limitation, space is also
circularly limited, in other words, if the spatial region of
concentration or limitation~$\mathcal{R}$ is a circle of radius~$R$,
then a polar coordinate, $\bx=(r,\theta)$, representation 
\be g(r,\theta)=\left\{
\begin{array}{l@{\quad\mbox{if}\hspace{0.6em}}l}
\rule[-2mm]{0mm}{6mm}\sqrt{2}\,g(r)\cos m\theta & m<0,\\
\rule[-2mm]{0mm}{6mm}g(r)                     & m=0,\\
\rule[-2mm]{0mm}{6mm}\sqrt{2}\,g(r)\sin m\theta & m>0,\\
\end{array}
\right.
\label{Bpolarg}
\ee
may be used to decompose eq.~(\ref{Dscaled})
into a series of non-degenerate fixed-order eigenvalue problems, after scaling,
\begin{subequations}
\label{Dscaled2}
\be
\label{Bscaled2}
\int_{0}^{1}D(\xi,\xi')\hsp\psi(\xi')\,
\xi'\hspace{0.05em}d\xi'=\lambda\hsp\psi(\xi),
\ee
\be
D(\xi,\xi')=4N\int_{0}^{1}\!
J_m\big(2\sqrt{N\DD}\,p\hspace{0.05em}\xi\big)\,
J_m\big(2\sqrt{N\DD}\,p\hspace{0.05em}\xi'\big)\,p\hsp dp
.
\ee
\end{subequations}
The solutions to eq.~(\ref{Dscaled2}) also solve a Sturm-Liouville
equation on $0\le\xi \le 1$. In terms of
$\varphi(\xi)=\sqrt{\xi}\,\psi(\xi)$, 
\be
\fracd{d}{d\xi}\left[(1-\xi^2)\frac{d\varphi}{d\xi}\right] + \left( \chi+
\fracd{1/4-m^2}{\xi^2} -4N\DD \xi^2  \right)\varphi =0
\label{gpsf},
\ee
for some~$\chi\ne\lambda$. When~$m=\pm1/2$ eq.~(\ref{gpsf}) reduces to
eq.~(\ref{slepian7.5}). By extension to~$\xi>1$
the fixed-order ``generalized prolate spheroidal functions''
$\varphi_1(\xi),\varphi_2(\xi),\dots$ can be determined from the
rapidly converging infinite series
\be
\varphi(\xi)=\frac{m!}{\gamma}\sum_{l=0}^\infty  
\frac{d_l\,l!}{(l+m)!}
\frac{J_{m+2l+1}(c\hsp\xi)}{\sqrt{c\hsp\xi}},\qquad \xi\in\mathbb{R},
\label{SE}
\ee
where the fixed-$m$ expansion coefficients~$d_l$ are determined by
recursion\cite{Slepian64} or  by diagonalization of a non-symmetric
tridiagonal matrix \cite{DeVilliers+2003,Shkolnisky2007} with elements
given by
\begin{eqnarray}
T_{l+1\,l}&=&-\fracd{c^2\,(m+l+1)^2}{(2l+m+1)(2l+m+2)},\nnr\\
T_{ll}&=&\left( 2l + m + \frac{1}{2}\right) \left(
2l+m+\frac{3}{2} \right)+\fracd{c^2}{2} \left[1+
\fracd{m^2}{(2l+m)(2l+m+2)} \right],\nnr\\
T_{l\,l+1}&=&-\fracd{c^2\,(l+1)^2}{(2l+m+2)(2l+m+3)},
\end{eqnarray}
where the parameter~$l$ ranges from~$0$ to
some large value that ensures convergence. The desired
concentration eigenvalues~$\lambda$ can subsequently be 
obtained by direct integration of eq.~(\ref{Dscaled}), or,
alternatively, from 
\be
\lambda=2\gamma^2\sqrt{N\DD},\with\gamma=
\frac{c^{m+1/2}d_0}{2^{m+1}(m+1)!}
\left(\sum_{l=0}^\infty d_l\right)^{-1} 
.
\ee
Numerous numerical methods exist to use
eqs~(\ref{Beigen12})--(\ref{Beigen34}) and~(\ref{gpsf}) in solving the
concentration problem~(\ref{Bnormratio}) for a variety of
two-dimensional Cartesian domains. After a long hiatus since the work
of Slepian\cite{Slepian64}, the two-dimensional problem has recently
been the focus of renewed attention in the applied mathematics
community\cite{DeVilliers+2003,Shkolnisky2007}, and applications
to the geosciences are to follow\cite{Simons+2010} .

\begin{figure}\center
\includegraphics[width=0.9\columnwidth]{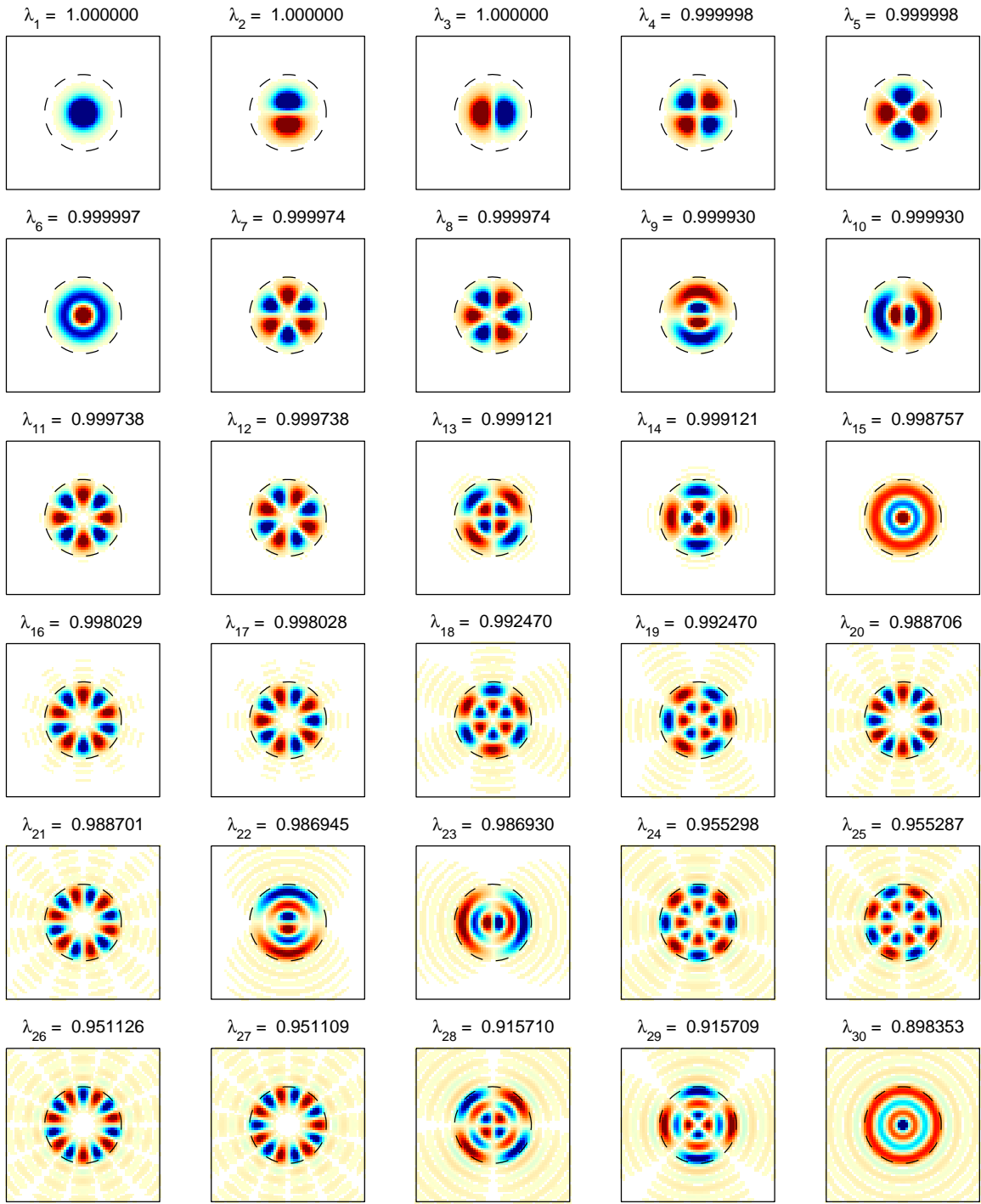}
\caption{\label{swfried2d}Bandlimited eigenfunctions~$g_\alpha(r,\theta)$
that are optimally concentrated within a Cartesian disk of
radius~$R=1$. The dashed circle denotes the region boundary. The Shannon
number~$N\DD=42$. The eigenvalues $\lambda_{\alpha}$ have been
sorted to a global ranking with the best concentrated eigenfunction
plotted at the top left and the $30$th best in the lower right. Blue
is positive and red is negative and the color axis is symmetric, but
the sign is arbitrary; regions in which the absolute value is less
than one hundredth of the maximum value on the domain are left
white. The calculations were performed by Gauss-Legendre integration
in the two-dimensional spatial domain, which sometimes leads to slight
differences in the last two digits of what should be identical
eigenvalues for each pair of non-circularly-symmetric eigenfunctions.}
\end{figure}

An example of Slepian functions on a disk-shaped region
in the Cartesian plane can be found in Figure~\ref{swfried2d}.

\ssec{Spatiospectral concentration on the surface of a sphere}

\ssec*{General theory in ``three'' dimensions}

We denote the colatitude of a geographical point~$\rhat$ on the
unit sphere surface $\Omega=\{\rhat: \|\rhat\|=1\}$ by
$0\le\theta\le\pi$ and the longitude by $0\le\phi< 2\pi$. We use~$R$
to denote a region of~$\Omega$, of area~$A$, within which we seek to
concentrate a bandlimited function of position~$\rhat$. We express a
square-integrable function~$f(\rhat)$ on the surface of the unit
sphere as 
\be
\label{expansion}
f(\rhat)=\sumsh {f}_{lm}\Ylmrh,\qquad {f}_{lm}=\into
f(\rhat)\hspace*{0.05em}\Ylmrh\domg,
\ee
using orthonormalized real surface spherical
harmonics~\cite{Edmonds96,Dahlen+98} 
\ber
\Ylmrh= \Ylm(\theta,\phi)&=&\left\{
\begin{array}{l@{\quad\mbox{if}\hspace{0.6em}}l}
\rule[-2mm]{0mm}{6mm}\sqrt{2}X_{lm}(\theta)\cos m\phi & -l\le m<0,\\
\rule[-2mm]{0mm}{6mm}X_{l0}(\theta)                     & m=0,\\
\rule[-2mm]{0mm}{6mm}\sqrt{2}\Xlm(\theta)\sin m\phi & 0< m\le l,\\
\end{array}
\right.\label{Ylm}\\
\Xlm(\theta)&=&
(-1)^m\tlofp^{1/2}
\left[\frac{(l-m)!}{(l+m)!}\right]^{1/2}\!\Plm (\cos\theta),
\label{xlm}\\
\Plm (\mu)&=&
\frac{1}{2^ll!}(1-\mu^2)^{m/2}\left(\frac{d}{d\mu}\right)^{l+m}\!(\mu^2-1)^l.
\label{plm}
\eer
The quantity $0\le l\le\infty$ is the angular degree of the
spherical harmonic, and $-l\le m\le l$ is its angular order. The
function~$\Plm(\mu)$ defined in (\ref{plm}) is the associated
Legendre function of integer degree~$l$ and order~$m$. Our
choice of the constants in eqs~(\ref{Ylm})--(\ref{xlm})
orthonormalizes the harmonics on the unit sphere:
\be
\into\Ylm\Ylmp\domg=\dllp\dmmp
,
\label{normalization}
\ee
and leads to the addition theorem in terms of the Legendre
functions~$P_l(\mu)=P_{l0}(\mu)$ as
\be
\suml_{m=-l}^l\Ylmrh\Ylmrhp =
\tlofp P_l(\rhat\cdot\rhat').
\label{additionSH}
\ee
To maximize the spatial concentration of a bandlimited
function
\be
g(\rhat)=\sumshL g_{lm}\Ylmrh
\label{bandlg}
\ee
within a region~$R$, we maximize the energy ratio
\be
\lambda=\fracd{\intr g^2(\rhat)\domg}{\into^{}g^2(\rhat)\domg}=\mbox{maximum}
.
\label{normratio}
\ee
Maximizing equation~(\ref{normratio}) leads to the 
positive-definite spectral-domain eigenvalue equation
\begin{subequations}
\label{fulleigen1}
\be
\label{halfeigen}
\suml_{l'=0}^L\suml_{m'=-l'}^{l'}\Dlmlmp
g_{l'm'}=\lambda\hspace{0.05em}g_{lm},\qquad 0\le l\le L,
\ee
\be
\Dlmlmp=\intr\Ylm\Ylmp\domg,
\label{Dlmlmpdef}
\ee
\end{subequations}
and we may equally well rewrite eq.~(\ref{fulleigen1}) as a
spatial-domain eigenvalue equation: 
\begin{subequations}
\label{firsttimeint}
\be
\label{eig3}
\intr  \Drhrhp \,\grhp\domg'=
\lambda\hspace{0.05em}g(\rhat),\qquad \rhat\in\Omega,
\ee
\be
\Drhrhp
=\sum_{l=0}^L\tlofp\!P_l(\rhat\cdot\rhat'),
\label{banddelta}
\ee
\end{subequations}
where~$P_l$ is the Legendre function of
degree~$l$. Eq.~(\ref{firsttimeint}) is a homogeneous Fredholm
integral equation of the second kind, with a finite-rank, symmetric,
Hermitian kernel. We choose the spectral eigenfunctions of the operator in
eq.~(\ref{Dlmlmpdef}), whose elements are $\glma$, $\alpha=1,$
$\dots,$ $\Lpot$, to satisfy the orthonormality relations 
\be
\label{orthogn}
\sumshortL\galm\glmb=\dab,\qquad
\sumshortL\galm\sumshortLp\Dlmlmp\gblmp=\lambda_\alpha\dab
.
\ee
The finite set of bandlimited spatial eigensolutions
$g_1(\rhat),g_2(\rhat), \ldots,g_{(L+1)^2}(\rhat)$ can be made
orthonormal over the whole sphere~$\Omega$ and orthogonal over the
region~$R$:  
\be
\into g_{\alpha}g_{\beta}\domg=\delta_{\alpha\beta},\qquad
\intr  g_{\alpha}g_{\beta}\domg=\lambda_{\alpha}\delta_{\alpha\beta}.
\label{orthog}
\ee
In the limit of a small area $A\rightarrow 0$ and a large bandwidth 
$L\rightarrow\infty$ and after a change of variables, a scaled version
of eq.~(\ref{firsttimeint}) will be given by 
\begin{subequations}
\label{3Dscaled}
\be
\int_{R_{*}}\!\!\Dxixip
\,\psi(\bxi')\,d\Omega'_*=\lambda\hsp\psi(\bxi),
\ee
\be
\Dxixip=\frac{(L+1)\sqrt{A/4\pi}}{2\pi}
\fracd{J_1[(L+1)\sqrt{A/4\pi}\,\,\|\bxi-\bxi'\|]} 
{\|\bxi-\bxi'\|},
\ee
\end{subequations}
where the scaled region~$R_{*}$  now has area~$4\pi$ and~$J_1$ again
is the first-order Bessel function of the first kind. As in the
one- and two-dimensional case, the asymptotic, or ``flat-Earth''
eigenvalues $\lambda_1\ge\lambda_2\ge\ldots$ and scaled eigenfunctions
$\psi_1(\bxi),\psi_2(\bxi),\ldots$ depend upon the maximal
degree~$L$ and the area~$A$ only through what is once again a
space-bandwidth product, the ``spherical Shannon number'',
this time given by  
\ber
\nnr
N\DDD&=&\sum_{\alpha =1}^{\Lpot}\lambda_{\alpha}=
\sum_{l=0}^L\sum_{m=-l}^l 
D_{lm,lm}=\int_RD(\rhat,\rhat)\,d\Omega\\
&=&\int_{R_{*}}\!\!\Dxixi
\,d\Omega_*
=\Lpot\,\frac{A}{4\pi}
\label{tracedef}
.
\eer
Irrespectively of the particular region of concentration that they
were designed for, the complete set of bandlimited spatial Slepian
eigenfunctions $g_1,g_2, \ldots,g_{(L+1)^2}$, is a basis for
bandlimited scalar processes anywhere on the surface of the unit
sphere\cite{Simons+2006a,Simons+2006b}. This follows directly from the
fact that the spectral localization kernel~(\ref{Dlmlmpdef}) is real,
symmetric, and positive definite: its eigenvectors $g_{1\hsp lm},
g_{2\hsp lm}, \ldots, g_{\Lpot \hsp lm}$ form an orthogonal set as we
have seen. Thus the Slepian basis functions $\garh$,
$\alpha=1,\ldots,\Lpot$ given by eq.~(\ref{bandlg}) simply transform
the same-sized limited set of spherical
harmonics~$\Ylmrh$, $0\le l\le L$, $-l\le m\le l$ that are a basis
for the same space of bandlimited 
spherical functions with no power above the bandwidth~$L$. The effect
of this transformation is to order the resulting basis set such that
the energy of the first~$N\DDD$ functions, $g_{1}(\rhat),\ldots,
g_{N\DDD}(\rhat)$, with eigenvalues $\lambda\approx 1$, are
concentrated in the region~$R$, whereas the remaining eigenfunctions,
$g_{N\DDD+1}(\rhat),\ldots,g_{\Lpot}(\rhat)$, are concentrated in the
complimentary region~$\Omega-R$. As in the one- and two-dimensional
case, therefore, the reduced set of basis functions
$g_1,g_2,\ldots,g_{N\DDD}$ can be regarded as a sparse, global, basis
suitable to approximate bandlimited processes that are primarily
localized to the region~$R$. The dimensionality reduction is dependent
on the fractional area of the region of interest. In other words, the
full dimension of the space~$\Lpot$ can be ``sparsified'' to an effective
dimension of $N\DDD=\Lpot A/(4\pi)$ when the signal of interest
resides in a particular geographic region.

\begin{figure}[bht]\center
\rotatebox{-90}{\includegraphics[height=0.8\columnwidth]
{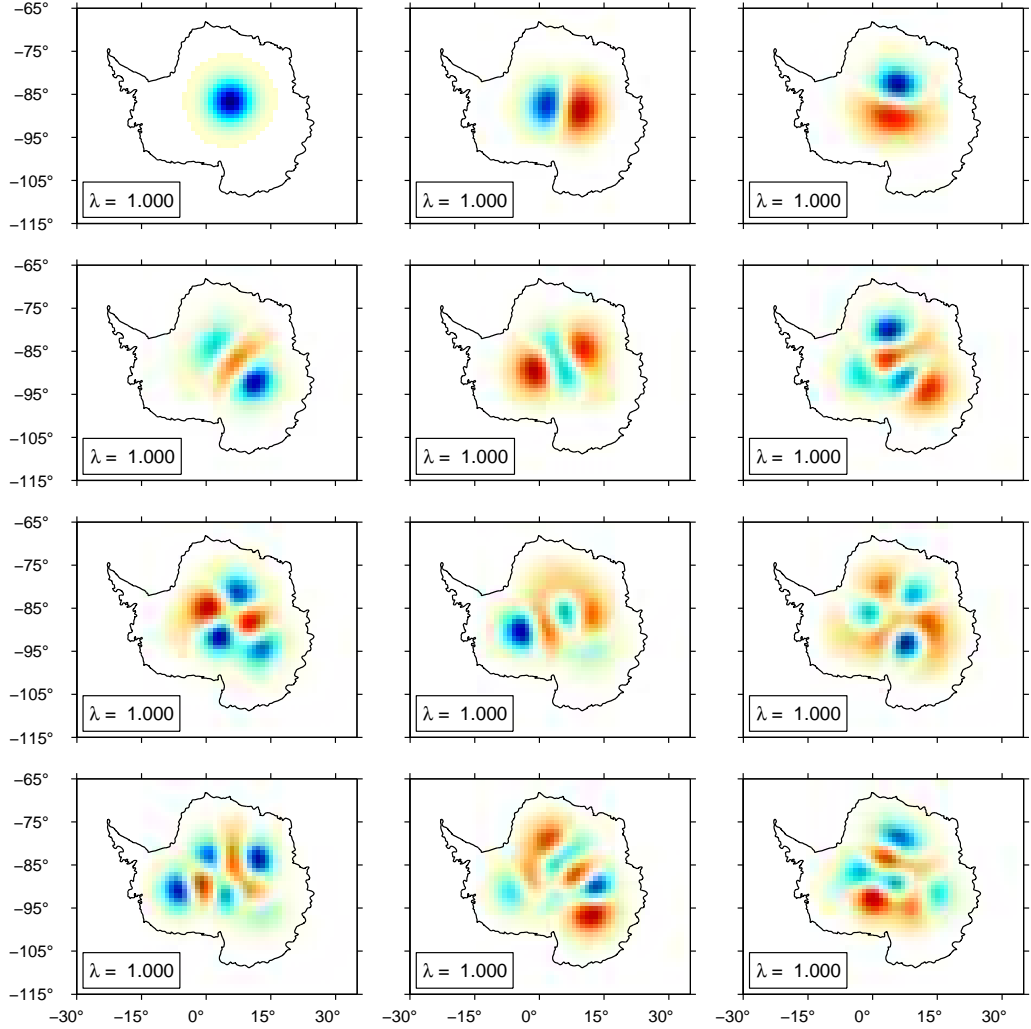}}
\caption{\label{antarctica}
Bandlimited $L=60$ eigenfunctions
$g_1,g_2,\ldots,g_{12}$ that are optimally concentrated within
Antarctica. The concentration factors
$\lambda_1,\lambda_2,\ldots,\lambda_{12}$ are indicated; the rounded
Shannon number is $N\DDD=102$.  The order of concentration is left to right,
top to bottom. Positive values are blue and negative values are red;
the sign of an eigenfunction is arbitrary. Regions in which the
absolute value is less than one hundredth of the maximum value on the
sphere are left white. We integrated eq.~(\ref{Dlmlmpdef}) over a
splined high-resolution representation of the boundary, using
Gauss-Legendre quadrature over the colatitudes, and analytically in
the longitudinal dimension\cite{Simons+2007}. 
}
\end{figure}

An example of Slepian functions on a geographical domain on the
surface of the sphere is found in Figure~\ref{antarctica}.  

\sssec*{Sturm-Liouville character and tridiagonal matrix formulation}

In the special but important case in which the region of concentration
is a circularly symmetric cap of colatitudinal radius~$\Theta$,
centered on the North Pole, the colatitudinal parts~$g(\theta)$ of the
separable functions  
\be
g(\theta,\phi)=\left\{
\barray{l@{\quad\mbox{if}\hspace{0.6em}}l}
\rule[-2mm]{0mm}{6mm}\sqrt{2}\,g(\theta)\cos m\phi & -L\le m<0,\\
\rule[-2mm]{0mm}{6mm}g(\theta)                     & m=0,\\
\rule[-2mm]{0mm}{6mm}\sqrt{2}\,g(\theta)\sin m\phi & 0< m\le L,\\
\earray
\right.
\label{polarg2}
\ee
which solve eq.~(\ref{firsttimeint}), or, indeed, the fixed-order
versions 
\begin{subequations}
\be
\int_{0}^{\Theta}D(\theta,\theta')\,g(\theta')\sin\theta'\,d\theta'
={\lambda}\hspace{0.05em}g(\theta),\quad 0\le\theta\le\pi,
\label{Fredholm}
\ee
\be
D(\theta,\theta')=2\pi\suml_{l=m}^{L}\Xlmth \Xlmthp,
\label{dlx}
\ee
\end{subequations}
are identical to those of a Sturm-Liouville equation for
the~$g(\theta)$. In terms of $\mu=\cos\theta$, 
\be
\fracd{d}{d\mu}\left[(\mu-\cos\Theta)
(1-\mu^2)\fracd{dg}{d\mu}\right]+
\left(\chi+
L(L+2)\mu
-\fracd{m^2(\mu-\cos\Theta)}{1-\mu^2}
\right)g=0,
\label{grundiffeqn}
\ee
with~$\chi\ne\lambda$. This equation can be solved in the spectral
domain by diagonalization of a simple tridiagonal matrix with a
very well-behaved spectrum\cite{Simons+2006a,Simons+2007}. This
matrix, whose eigenfunctions correspond to the~$g_{lm}$ of
eq.~(\ref{bandlg}) at  constant~$m$ is given by
\ber T_{ll}&=&-l(l+1)\cos{\Theta},\nnr\\
T_{l\,l+1}&=&\big[l(l+2)-L(L+2)\big]
\sqrt{\fracd{(l+1)^2-m^2}{(2l+1)(2l+3)}},
\label{gdefi1}
\eer
Moreover, when the region of concentration is a \textit{pair} of axisymmetric
polar caps of common colatitudinal radius~$\Theta$ centered on the
North and South Pole, the~$g(\theta)$ can be obtained by
solving the Sturm-Liouville equation
\be
\fracd{d}{d\mu}\left[(\mu^2-\cos^2\Theta)
(1-\mu^2)\fracd{dg}{d\mu}\right]
+\left(
\chi + L_p(L_p+3)\mu^2
-\fracd{m^2(\mu^2-\cos^2\Theta)}{1-\mu^2}
\right)g=0,\label{grundiffeqn2}
\ee
where~$L_p=L$ or~$L_p=L-1$ depending whether the order~$m$ of the
functions~$g(\theta,\phi)$ in eq.~(\ref{polarg2}) is odd or
even, and whether the bandwidth~$L$ itself is odd or even. In their
spectral form the coefficients of the optimally concentrated antipodal
polar-cap eigenfunctions only require the numerical diagonalization of
a tridiagonal matrix with analytically prescribed elements and a
spectrum of eigenvalues that is guaranteed to be
simple\cite{Simons+2006b,Simons+2007}, namely 
\ber
T^p_{ll}&=&-l(l+1)\cos^2{\Theta}
+\frac{2}{2l+3}\left[(l+1)^2-m^2\right]\nnr\\
&&{}+[(l-2)(l+1)-L_p(L_p+3)]
\left[\frac{1}{3}-\frac{2}{3}\,
\fracd{3m^2-l(l+1)}{(2l+3)(2l-1)}\right],\nnr\\ 
T^p_{l\,l+2}&=&\fracd{\big[l(l+3)-L_p(L_p+3)\big]}{2l+3}
\sqrt{\fracd{\left[(l+2)^2-m^2\right]
\left[(l+1)^2-m^2\right]}{(2l+5)(2l+1)}}.
\label{gdefi}
\eer
The concentration values~$\lambda$, in turn, can be determined from
the defining equations~(\ref{fulleigen1}) or~(\ref{firsttimeint}). The
spectra of the eigenvalues~$\chi$ of eqs~(\ref{gdefi1})
and~(\ref{gdefi}) display roughly equant spacing, without the
numerically troublesome plateaus of nearly equal values that
characterizes the eigenvalues~$\lambda$. Thus, for the special cases
of symmetric single and double polar caps, the concentration problem
posed in eq.~(\ref{normratio}) is not only numerically feasible also
in circumstances where direct solution methods are bound to
fail\cite{Albertella+99}, but essentially trivial in every
situation. Numerical methods for the solution
of eqs~(\ref{fulleigen1})--(\ref{firsttimeint}) on completely general
domains on the surface of the sphere were discussed by us
elsewhere\cite{Simons+2006a,Simons+2006b,Simons+2007}.

\ssec{Mid-term summary}

It is interesting to reflect, however heuristically, on the commonality of
all of the above aspects of spatiospectral localization, in the
slightly expanded context of reproducing-kernel Hilbert
spaces\cite{Nashed+91,Daubechies92,Yao67,Amirbekyan+2008b}. In one  
dimension, the Fourier orthonormality relation and the ``reproducing''
properties of the spatial delta function are given by 
\be 
\label{BFortho0}
\delta(t,t')=
\jnorm\intinf
e^{i\omega(t-t')}\,d\omega
,\qquad \intinf f(t')\hsp\delta(t,t')\,dt'=f(t)
.
\ee
In two Cartesian dimensions the equivalent relations are
\be
\label{BFortho00}
\delta(\bx,\bx')=\fnorm\intinf
e^{i\bk\cdot(\bx-\bx')}\dbk
,\qquad \intinf f(\bx')\hsp\delta(\bx,\bx')\dbx'=f(\bx)
,
\ee
and on the surface of the unit sphere we have
\be
\label{BFortho000}
\delta(\rhat,\rhat')=\sum_{l=0}^\infty\tlofp\!P_l(\rhat\cdot\rhat')
,\qquad
\into f(\rhat')\hsp\delta(\rhat,\rhat')\domg'=f(\rhat)
.
\ee
The integral-equation kernels~(\ref{slepian4}), (\ref{Beigen4})
and~(\ref{banddelta}) are all bandlimited spatial delta functions
which are reproducing kernels for bandlimited functions of the
types in eqs~(\ref{gband}), (\ref{Bgdefn}) and~(\ref{bandlg}): 
\ber
\label{dttp}
\Dttp=\jnorm\intW 
e^{i\omega(t-t')}\,d\omega,\qquad&&\intinf g(t')\Dttp\,dt'=g(t),\\
\label{dxxp}
\Dxxp=
\fnorm
\intK e^{i\bk\cdot(\bx-\bx')}\dbk,\qquad&&\intinf
g(\bx')\Dxxp\dbx'=g(\bx),\\ 
\label{drrp}
\Drhrhp=\sum_{l=0}^L\tlofp\!P_l(\rhat\cdot\rhat'),\qquad&&\into
g(\rhat')\Drhrhp\domg=g(\rhat).
\eer
The equivalence of eq.~(\ref{dttp}) with eq.~(\ref{slepian4}) is
through the Euler identity and the reproducing properties follow from
the spectral forms of the orthogonality
relations~(\ref{BFortho0})--(\ref{BFortho00}), which are self-evident
by change-of-variables, and from the spectral form of
eq.~(\ref{BFortho000}), which is eq.~(\ref{normalization}). Much as
the delta functions of eqs~(\ref{BFortho0})--(\ref{BFortho000}) set up
the Hilbert spaces of all square-integrable functions on the real
line, in two-dimensional Cartesian space, and on the surface of the
sphere, the kernels~(\ref{dttp})--(\ref{drrp}) induce the equivalent
subspaces of bandlimited functions in their respective
dimensions. Inasmuch as the Slepian functions are the integral
eigenfunctions of these reproducing kernels in the sense of
eqs.~(\ref{eig1}), (\ref{eig2}) and~(\ref{eig3}), they are complete
bases for their band-\textit{limited}
subspaces\cite{Slepian+61,Landau+61,Daubechies92,Flandrin98,Freeden+98c}.
Therein, the~$N\D$, $N\DD$ or~$N\DDD$ best time- or
space-\textit{concentrated} members allow for sparse, approximate,
expansions of signals that are spatially concentrated to
the one-dimensional interval $t\in[-T, T]\subset\mathbb{R}$, the
Cartesian region $\bx \in\mathcal{R}\subset \mathbb{R}^2$, or the
spherical surface patch $\rhat\in R\subset\Omega$.

As a corollary to this behavior, the infinite sets of \textit{exactly}
time- or spacelimited (and thus band-\textit{concentrated}) versions of
the functions~$g$, which are the eigenfunctions of eqs~(\ref{dude1}),
(\ref{Beigen34}) and~(\ref{firsttimeint}) with the domains
appropriately restricted, are complete bases for square-integrable
scalar functions on the intervals to which they are
confined\cite{Slepian+61,Landau+61,Simons+2006a}. Expansions of
such wideband signals in the small subset of their 
$N\D$, $N\DD$ or~$N\DDD$ most band-concentrated members provide
reconstructions which are constructive in the sense that
they progressively capture all of the signal in the mean squared sense,
in the limit of letting their numbers grow to
infinity. This second class of functions can be trivially obtained, up
to a multiplicative constant, from the bandlimited Slepian
functions~$g$ by simple time- or space limitation. While
Slepian\cite{Slepian+61,Slepian83}, for this reason perhaps, never
gave them a name, we have been referring to those as~$h$ in our own
investigations of Slepian functions on the
sphere\cite{Simons+2006a,Simons+2006b,Dahlen+2008}. 





\section{Problems in the geosciences and beyond}

Taking all of the above at face-value but referring again to the
literature cited thus far for proof and additional context, we return 
to considerations closer to home, namely the estimation of geophysical
(or cosmological) signals and/or their power spectra, from noisy and
incomplete observations collected at or above the surface of the 
spheres ``Earth'' or ``planet'' (or from inside the sphere
``sky''). We restrict ourselves to real-valued scalar measurements, 
contaminated by additive noise for which we shall adopt idealized
models. We focus exclusively on data acquired and solutions
expressed on the \textit{unit} sphere. We have considered
generalizations to problems involving satellite data collected at an
altitude and/or potential fields
elsewhere\cite{Wieczorek+2005,Simons+2006b,Simons+2007,Dahlen+2008}. Two 
different statistical problems come up in this context, namely, (i)~how
to find the ``best'' estimate of the signal given the data, and
(ii)~how to construct from the data the ``best'' estimate of the power
spectral density of the signal in question.  

Thus, let there be some data distributed on the unit sphere,
consisting of ``signal'', $s$ and ``noise'', $n$, and let there be
some region of interest $R\subset\Omega$, in other words, let
\ber
d(\rhat)=\left\{\begin{array}{ll}
s(\rhat)+n(\rhat) & \mbox{if $\rhat\in R$},\\
\mbox{unknown/undesired} & \mbox{if $\rhat\in R-\Omega$}.
\end{array}\right.
\label{datdefs}
\eer
We assume that the signal of interest can be expressed by way of
spherical harmonic expansion as in eq.~(\ref{expansion}), and that it
is, itself, a realization of a zero-mean, Gaussian, isotropic, random
process, namely 
\be
s(\rhat)=\sumsh s_{lm}\Ylmrh, \qquad
\langle s_{lm}\rangle =0\also 
\langle s_{lm}s_{l'm'}\rangle=S_l\,\dllp\dmmp.
\ee
For illustration we furthermore assume that the noise is a zero-mean
stochastic process with an isotropic power spectrum, i.e. $\langle
n(\rhat)\rangle=0$ and $\langle
n_{lm}n_{l'm'}\rangle=N_l\,\dllp\dmmp$, and that it is 
statistically uncorrelated with the signal. We refer to 
power as \textit{white} when $S_l=S$ or $N_l=N$, or, equivalently,
when $\langle n(\rhat)n(\rhat')\rangle=N\delta(\rhat,\rhat')$. 
Our objectives are thus (i)~to determine the best
estimate~$\hat{s}_{lm}$ of the spherical harmonic expansion
coefficients~$s_{lm}$ of the signal and 
(ii)~to find the best estimate~$\hat{S}_l$ for the isotropic power
spectral density~$S_l$. While in the physical world there can be no limit
on bandwidth, practical restrictions force any and all of our
estimates to be bandlimited to some maximum spherical harmonic degree
$L$, thus by necessity $\hat{s}_{lm}=0$ and $\hat{S}_l=0$ for~$l> L$:
\be\label{wouldbe}
\hat{s}(\rhat)=\sumshL \hat{s}_{lm}\Ylmrh
.
\ee
This limitation, combined with the statements eq.~(\ref{datdefs}) on the
data coverage or the study region of interest, naturally leads us back
to the concept of ``spatiospectral concentration'', and, as we shall
see, solving either problem~(i) or~(ii) will gain from involving the
``localized'' Slepian functions rather than, or in addition to, the
``global'' spherical harmonics basis.

This leaves us to clarify what we understand by ``best'' in this
context. While we adopt the traditional statistical metrics of bias,
variance, and mean squared error to appraise the quality of our
solutions\cite{Cox+74,Bendat+2000}, the resulting connections to
sparsity will be real and immediate, owing to the Slepian functions
being naturally instrumental in constructing
efficient, consistent and/or unbiased estimates of
either~$\hat{s}_{lm}$ or~$\hat{S}_l$. Thus, we define
\be
v=\langle\shat^2\rangle-\langle\shat\rangle^2
,\qquad 
b=\langle\shat\rangle-s,\qquad 
\epsilon=\shat-s,
\also
\langle\epsilon^2\rangle=v+b^2
\ee
for problem~(i), where the lack of subscript indicates that we can
study variance, bias and mean squared error of the estimate of the
coefficients~$\hat{s}_{lm}$ but also of their spatial expansion
$\shat(\rhat)$. For problem~(ii) on the other hand, we focus 
on the estimate of the isotropic power spectrum at a given spherical
harmonic degree~$l$ by identifying 
\be
v_l=\langle\Shat^2_l\rangle-\langle\Shat_l\rangle^2,\qquad 
b_l=\langle\Shat_l\rangle-S_l,\qquad
\epsilon_l=\Shat_l-S_l,\also
\langle\epsilon^2_l\rangle=v_l+b^2_l
.
\ee
Depending on the application, the ``best'' estimate could mean the
unbiased one with the lowest
variance\cite{Tegmark97b,Tegmark+97,Bond+98,Oh+99,Hinshaw+2003}, it
could be simply the minimum-variance estimate having some acceptable
and quantifiable bias\cite{Wieczorek+2007}, or, as we would usually
prefer, it would be the one with the minimum mean squared
error\cite{Simons+2006b,Dahlen+2008}. 

\ssec{Problem~(i): Signal estimation from noisy and incomplete
  spherical data}

\sssec*{Spherical harmonic solution}

Paraphrasing results elaborated elsewhere\cite{Simons+2006b}, we 
write the bandlimited solution to the damped inverse
problem 
\be
\intr(s-d)^2\domg+\eta\intbr s^2\domg=\mbox{minimum}
,
\label{variational}
\ee
where~$\eta\ge 0$ is a damping parameter, by straightforward algebraic
manipulation, as 
\be
\hat{s}_{lm}=\sumshLp \left(\Dlmlmp+\eta\bDlmlmp\right)^{-1} 
\intr d\,\Ylmp\domg
\label{hatss}
,
\ee
where~$\bDlmlmp$, the kernel that localizes to the region~$\Omega-R$,
compliments~$\Dlmlmp$ given by eq.~(\ref{Dlmlmpdef}) which localizes
to~$R$. Given the eigenvalue spectrum of the latter, its inversion is
inherently unstable, thus eq.~(\ref{variational}) is an
ill-conditioned inverse problem unless~$\eta>0$, as has been well
known, e.g. in
geodesy\cite{Xu92a,Sneeuw+97}. Elsewhere\cite{Simons+2006b} we have 
derived exact expressions for the optimal value of the damping
parameter~$\eta$ as a function of the signal-to-noise ratio under
certain simplifying assumptions. As can be easily shown, without
damping the estimate is unbiased but effectively incomputable; the
introduction of the damping term stabilizes the solution at the cost
of added bias. And of course when~$R=\Omega$, eq.~(\ref{hatss}) is
simply \textit{the}  spherical harmonic transform, as in that case,
eq.~(\ref{Dlmlmpdef}) reduces to eq.~(\ref{normalization}), in other
words, then~$\Dlmlmp=\dllp\dmmp$.   

\sssec*{Slepian basis solution}

The trial solution in the Slepian basis designed for this
region of interest~$R$, i.e.
\be
\hat{s}(\rhat)=\sumapot\hat{s}_\alpha \garh 
,
\label{shatdef1}
\ee
would be completely equivalent to the expression in
eq.~(\ref{wouldbe}) by virtue of the completeness of the Slepian basis
for bandlimited functions everywhere on the sphere and the unitarity
of the transform~(\ref{bandlg}) from the spherical
harmonic to the Slepian basis. The solution to the undamped ($\eta=0$)
version of eq.~(\ref{variational}) would then be 
\be
\hat{s}_\alpha=\lambda_\alpha^{-1}\intr d\hsp g_\alpha\domg
,\label{SGsolspec}
\ee
which, being completely equivalent to eq.~(\ref{hatss}) for
$\eta=0$, would be computable, and biased, only when the expansion in
eq.~(\ref{shatdef1}) were to be truncated to some finite
$J<\Lpot$ to prevent the blowup of the eigenvalues~$\lambda$. Assuming
for simplicity of the argument that~$J=N\DDD$, the 
essence of the approach is now that the solution 
\be
\hat{s}(\rhat)=\sumaN\hat{s}_\alpha \garh
\label{shatdef2}
\ee
will be sparse (in achieving a bandwidth~$L$ using~$N\DDD$
Slepian instead of~$\Lpot$ spherical-harmonic expansion coefficients)
yet good (in approximating the signal as well as possible in the mean
squared sense in the region of interest~$R$) and of geophysical
utility (assuming we are dealing with spatially localized processes
that are to be extracted, e.g., from global satellite
measurements)\cite{Han+2008a,Simons+2009}.    

\ssec*{Bias and variance}

In concluding this section let us illustrate another welcome
by-product of our methodology, by writing the mean squared error for
the spherical harmonic solution~(\ref{hatss}) compared to the
equivalent expression for the solution in the Slepian basis,
eq.~(\ref{SGsolspec}). We do this as a function of the spatial
coordinate, in the Slepian basis for both, and, for maximum clarity of
the exposition, using the contrived case when both signal and noise
should be white as well as bandlimited (which technically is
impossible). In the former case, we 
get
\ber
\label{msefinal}
\langle\epsilon^2(\rhat)\rangle&=&N\sumapot\lambda_\alpha
[\lambda_\alpha+\eta(1-\lambda_\alpha)]^{-2}
g_\alpha^2(\rhat)\\
&&{}+\eta^2 S\sumapot
(1-\lambda_\alpha)^2[\lambda_\alpha+\eta(1-\lambda_\alpha)]^{-2}
g_\alpha^2(\rhat)\nnr 
,
\eer
while in the latter, we obtain
\be
\langle\epsilon^2(\rhat)\rangle=N\sumaN\lambda_\alpha^{-1}g_\alpha^2(\rhat)+S\sumakR
g_\alpha^2(\rhat)
.
\label{SGmsefinal}
\ee
All~$\Lpot$ basis functions are required to express the mean squared
estimation error, whether in eq.~(\ref{msefinal}) or in
eq.~(\ref{SGmsefinal}). The first term in both expressions is the
variance, which depends on the measurement noise. Without damping or truncation
the variance grows without bounds. Damping and truncation alleviate
this at the expense of added bias, which depends on the
characteristics of the signal, as given by the second term. In contrast
to eq.~(\ref{msefinal}), however, the Slepian
expression~(\ref{SGmsefinal}) has disentangled the contributions due
to noise/variance and signal/bias by projecting them onto the sparse
set of well-localized and the remaining set of poorly localized Slepian
functions, respectively. The estimation variance is felt via the
basis functions $\alpha=1\rar N\DDD$ that are well concentrated inside
the measurement area, and the effect of the bias is relegated to those
$\alpha=N\DDD+1\rar\Lpot$ functions that are confined to the region of
missing data. 

When forming a solution to problem~(i) in the Slepian basis by
truncation according to~eq.~(\ref{shatdef2}), changing the truncation
level to values lower or higher than the Shannon number~$N\DDD$
amounts to navigating the trade-off space between variance, bias (or
``resolution''), and sparsity in a manner that is captured with great
clarity by eq.~(\ref{SGmsefinal}). We refer the reader
elsewhere\cite{Simons+2006b,Simons+2007} for more details, and, in particular, for
the case of potential fields estimated from data collected at
satellite altitude.

\ssec{Problem~(ii): Power spectrum estimation from noisy and
incomplete spherical data}

Following \cite{Dahlen+2008} we will find it convenient to
regard the data~$d(\rhat)$ given in eq.~(\ref{datdefs}) as having been
multiplied by a unit-valued boxcar window function confined to the
region~$R$,   
\be
\label{boxcar}
b(\rhat)=\sum_{p=0}^\infty\sum_{q=-p}^pb_{pq}Y_{pq}(\rhat)
=\left\{\begin{array}{ll} 
1 & \mbox{if~$\rhat\in R$},\\
0 & \mbox{otherwise,} \end{array}\right.
\ee
The power spectrum of the boxcar window~(\ref{boxcar}) is 
\be \label{boxspec}
B_p=\frac{1}{2p+1}\sum_{q=-p}^pb_{pq}^2.
\ee

\sssec*{The spherical periodogram}

Should we decide that an acceptable estimate of the power spectral
density of the available data is nothing else but the weighted average of
its spherical harmonic expansion coefficients, we would be forming the
spherical analogue of what Schuster\cite{Schuster1898} named the
``periodogram'' in the context of time series analysis, namely
\be \label{SestSP} \hat{S}_l\SP=
\left(\frac{4\pi}{A}\right)\frac{1}{2l+1}\sum_{m=-l}^l
\left[\int_Rd(\rhat)\,\Ylmrh\domg
\right]^2
.
\ee

\sssec*{Bias of the periodogram}

Upon doing so we would discover that the expected value of
such an estimator would be the biased quantity
\be
\label{SexpecSP}
\langle \hat{S}_l\SP\rangle
=\sum_{l'=0}^\infty K_{ll'}(S_{l'}+N_{l'}),
\ee
where, as it is known in astrophysics and
cosmology\cite{Peebles73,Hauser+73,Hivon+2002}, the
periodogram ``coupling'' matrix
\be \label{Kmatdef}
K_{ll'}=\left(\frac{4\pi}{A}\right)\frac{1}{2l+1}
\sum_{m=-l}^l\sum_{m'=-l'}^{l'} 
\left[D_{lm,l'm'}\right]^2
,
\ee
governs the extent to which an estimate $\hat{S}_l\SP$ of~$S_l$ is
influenced by  spectral leakage from  power in
neighboring spherical harmonic degrees $l'=l\pm 1,l\pm 2,\ldots$, all
the way down to 0 and up to~$\infty$. In
the case of full data coverage, $R=\Omega$, or of a perfectly white
spectrum, $S_l=S$, however, the estimate would be unbiased ---
provided the noise spectrum, if known, can be subtracted beforehand. 

\sssec*{Variance of the periodogram}

The covariance of the periodogram estimator~(\ref{SestSP}) would moreover be
suffering from strong wideband coupling of the power spectral
densities in being given by  
\be \label{SigmaSP2}
\Sigma_{ll'}\SP=\frac{2(4\pi/A)^2}{(2l+1)(2l'+1)}\sum_{m=-l}^l\sum_{m'=-l'}^{l'}
\left[\sum_{p=0}^{\infty}\sum_{q=0}^{\infty}(S_p+N_p)
D_{lm,pq}D_{pq,l'm'}\right]^2, 
\ee
Even under the commonly made assumption as should the power spectrum
be slowly varying within the main lobe of the coupling matrix, such
coupling would be nefarious. In the ``locally white'' case we would have
\be \label{moderSP}
\Sigma_{ll'}\SP=\frac{2(4\pi/A)^2}{(2l+1)(2l'+1)}
\,(S_l+N_l)(S_{l'}+N_{l'})\sum_{m=-l}^l\sum_{m'=-l'}^{l'}
\left[D_{lm,l'm'}\right]^2.
\ee
Only in the limit of whole-sphere data coverage will
eqs~(\ref{SigmaSP2}) or~(\ref{moderSP}) reduce to  
\be \label{SigmaWSfin}
\Sigma_{ll'}\WS=\frac{2}{2l+1}\left(S_l+N_l\right)^2\dllp,
\ee
which is the ``planetary'' or  ``cosmic'' variance that can be
understood on the basis of elementary statistical considerations  
\cite{Jones63,Knox95,Grishchuk+97}. The strong spectral leakage for
small regions ($A\ll 4\pi$) is highly  undesirable and makes the
periodogram `hopelessly obsolete'\cite{Thomson+91}, or, to put it
kindly, `naive'\cite{Percival+93}, just as it is for one-dimensional
time series.

In principle it is possible  --- after subtraction of the noise
bias --- to eliminate the leakage bias in the
periodogram estimate~(\ref{SestSP}) by numerical inversion of the
coupling matrix~$K_{ll'}$. Such a 
``deconvolved periodogram'' estimator is unbiased. 
However, its covariance depends on the inverse of the periodogram
coupling matrix, which is only feasible when the region~$R$
covers most of the sphere, $A\approx 4\pi$. For any region whose area
$A$ is significantly smaller than~$4\pi$, the periodogram coupling
matrix~(\ref{Kmatdef}) will be too ill-conditioned to be invertible.

Thus, much like in problem~(i) we are faced with bad bias and poor
variance, both of which are controlled by the lack of localization of
the spherical harmonics and their non-orthogonality over incomplete
subdomains of the unit sphere. Both effects are described by the
spatiospectral localization kernel defined in~(\ref{Dlmlmpdef}),
which, in the quadratic estimation problem~(ii) appears in ``squared''
form in eq.~(\ref{SigmaSP2}). Undoing the effects of the wideband
coupling between degrees at which we seek to estimate the power
spectral density by inversion of the coupling kernel is virtually
impossible, and even if we could accomplish this to remove the
estimation bias, this would much inflate the estimation
variance\cite{Dahlen+2008}. 

\sssec*{The spherical multitaper estimate}

We therefore take a page out of the one-dimensional power estimation
playbook of Thomson\cite{Thomson82} by forming the
``eigenvalue-weighted multitaper
estimate''.  We could weight single-taper estimates adaptively to
minimize quality measures such as estimation variance or mean squared
error \cite{Thomson82,Wieczorek+2007}, but in practice, these methods
tend to be rather computationally demanding. Instead we simply
multiply the data~$d(\rhat)$ by the Slepian functions or
``tapers''~$g_{\alpha}(\rhat)$ designed for the region of interest
prior to computing power and then averaging:   
\be \label{SMTdef}
\hat{S}_l\MT=\sum_{\alpha=1}^{\Lpot}\lambda_\alpha
\left(\frac{4\pi}{N\DDD}\right)\frac{1}{2l+1}
\sum_{m=-l}^l\left[\int_{\Omega} g_{\alpha}(\rhat)\,
d(\rhat)\,Y_{lm}(\rhat)\domg\right]^2.
\ee

\sssec*{Bias of the multitaper estimate}

The expected value of the estimate~(\ref{SMTdef}) is
\be \label{SMTexpec}
\langle\hat{S}_l\MT\rangle=\sum_{l'=l-L}^{l+L}M_{ll'}(S_{l'}+N_{l'})
,
\ee
where the eigenvalue-weighted multitaper coupling matrix,
using Wigner 3-$j$ functions\cite{Varshalovich+88,Messiah2000}, is
given by
\be \label{MTcouple}
M_{ll'}=\frac{2l'+1}{\Lpot}\sum_{p=0}^L(2p+1)
\!\left(\!\begin{array}{ccc}
l & p & l' \\ 0 & 0 & 0\end{array}\!\right)^2.
\ee
It is remarkable that this result depends only upon the chosen bandwidth~$L$
and is completely independent of the size, shape or connectivity of
the region~$R$, even as~$R=\Omega$. Moreover, every row of the matrix
in eq.~(\ref{MTcouple}) sums to unity, which ensures that a
the multitaper spectral estimate $\hat{S}_l\MT$ has no leakage
bias in the case of a perfectly white spectrum provided the noise bias
is subtracted as well: $\langle\hat{S}_l\MT\rangle-\sum
M_{ll'}N_{l'}=S$ if~$S_l=S$.  

\sssec*{Variance of the multitaper estimate}

Under the moderately colored approximation, which is more easily
justified in this case because the coupling~(\ref{MTcouple}) is
confined to a narrow band of width 
less than or equal to~$2L+1$, with~$L$ the bandwidth of the tapers,
the eigenvalue-weighted multitaper covariance is 
\be \label{thisisit}
\Sigma_{ll'}\MT=\frac{1}{2\pi}(S_l+N_l)(S_{l'}+N_{l'})\sum_{p=0}^{2L}(2p+1)
\,\Gamma_p\!\left(\!\begin{array}{ccc}
l & p & l' \\ 0 & 0 & 0\end{array}\!\right)^2,
\ee
where, using Wigner 3-$j$ and 6-$j$
functions\cite{Varshalovich+88,Messiah2000}, 
\ber \label{Gammafin}
\Gamma_p&=&\frac{1}{(N\DDD)^2}\sum_{s=0}^L\sum_{s'=0}^L
\sum_{u=0}^L\sum_{u'=0}^L 
(2s+1)(2s'+1)(2u+1)(2u'+1)\nnr\\
&&{}\times\sum_{e=0}^{2L}(-1)^{p+e}(2e+1)B_e \hspace{8em}\nonumber\\
&&{}\times\left\{\!\begin{array}{ccc}
s & e & s' \\ u & p & u'\end{array}\!\right\}\left(\!\begin{array}{ccc}
s & e & s' \\ 0 & 0 & 0\end{array}\!\right)\!\left(\!\begin{array}{ccc}
u & e & u' \\ 0 & 0 & 0\end{array}\!\right)\!\left(\!\begin{array}{ccc}
s & p & u' \\ 0 & 0 & 0\end{array}\!\right)\!\left(\!\begin{array}{ccc}
u & p & s' \\ 0 & 0 & 0\end{array}\!\right).
\eer
In this expression $B_e$, the boxcar power~(\ref{boxspec}),  which we
note does depend on the shape of the region of interest, appears again,
summed over angular degrees limited by 3-$j$ selection rules to $0\leq
e\leq 2L$. The sum in eq.~(\ref{thisisit}) is likewise limited to
degrees $0\leq p\leq 2L$. The effect of tapering with windows
bandlimited to~$L$ is to introduce covariance between the estimates at
any two different degrees~$l$ and~$l'$ that are separated by fewer
than $2L+1$ degrees. Eqs~(\ref{thisisit})--(\ref{Gammafin}) are very
efficiently computable, which should make them competitive with, e.g.,
jackknifed estimates of the estimation variance
\cite{Chave+87,Thomson+91,Thomson2007}.  

The crux of the analysis lies in the fact that
the  matrix of the spectral covariances between
\textit{single-}tapered estimates is almost
diagonal\cite{Wieczorek+2007}, showing   
that the individual estimates that enter the weighted formula~(\ref{SMTdef})
are almost uncorrelated statistically. This embodies the very essence
of the multitaper method. It dramatically reduces the estimation
variance at the cost of small increases of readily quantifiable bias.

\section{Practical considerations}


In this section we now turn to the very practical context of sampled,
e.g. geodetic, data on the sphere. We shall deal exclusively with
bandlimited functions, which are equally well expressed in the
spherical harmonic as the Slepian basis, namely:  
\be\label{bandlf1} 
f(\rhat)=\sumshL f_{lm}\Ylmrh=\sumapot f_\alpha\hsp \garh
,
\ee
whereby the Slepian-basis expansion coefficients are obtained as
\be\label{bandlf2}
f_\alpha=\into f(\rhat)\garh\domg
.
\ee
If the function of interest is spatially localized in the region~$R$,
a truncated reconstruction using Slepian functions built for the
same region will constitute a very good, and sparse, local
approximation to it\cite{Simons+2009b}:  
\be\label{bandlfcon}
f(\rhat)\approx\sumaN f_\alpha\hsp \garh,\qquad\rhat\in R
.
\ee
We represent any sampled, bandlimited function $f$ by an
$M$-dimensional column vector  
\be\label{beff}
\beff=(f_1\; \cdots \; f_j \; \cdots\; f_M)\Tit
,
\ee
where $f_j=f(\br_j)$ is the value of $f$ at pixel   $j$,
and $M$ is the total number of sampling locations. In the most 
general case the distribution of pixel centers will be completely
arbitrary. The special case of equal-area pixelization of a 2-D
function $f(\br)$ on the unit sphere $\Omega$ is analogous to the
equispaced digitization of a  1-D time series. Integrals will then be
assumed to be approximated with sufficient accuracy by a Riemann sum
over a dense set of pixels,
\be \label{pixelsum}
\int\! f(\br)\domg\approx \Delta\Omega\sum_{j=1}^M f_j
\also
\int\! f^2(\br)\domg\approx\Delta\Omega\,\beff\T\beff
.
\ee
We have deliberately left the integration domain out of the above
equations to cover both the cases of sampling over the entire unit
sphere surface~$\Omega$, in which case the solid angle
$\Delta\Omega=4\pi/M$ (case~1) as well as over an incomplete
subdomain~$R\subset\Omega$, in which case $\Delta\Omega=A/M$, with $A$
the area of the region~$R$ (case~2).  If we collect  the real spherical
harmonic basis functions $\Ylm$ into an $\Lpot\times M$-dimensional matrix  
\be
\bY=
\left(\barray{ccccc}
Y_{00}(\rhat_1) & \cdots     & Y_{00}(\rhat_j)  & \cdots &  Y_{00}(\rhat_M)\\
                &            & \vdots           &        &       \\
\cdots          &            & \Ylm(\rhat_j)    &        &\cdots \\
                &            & \vdots           &        &       \\
Y_{LL}(\rhat_1) & \cdots     & Y_{LL}(\rhat_j)  & \cdots &Y_{LL}(\rhat_M)       
\earray
\right)
,
\ee
and the spherical harmonic coefficients of the function into an
$\Lpot\times 1$-dimensional vector 
\be
\label{fdef}
\fb=(\;f_{00}\;\cdots\; f_{lm}\; \cdots\; f_{LL}\;)\Tit
,
\ee
we can write the spherical harmonic synthesis in eq.~(\ref{bandlf1})
for sampled data without loss of generality as 
\be
\label{synth}
\beff=\bY\T \fb
.
\ee
We will adhere to the notation convention of using sans-serif fonts
(e.g. $\beff$, $\bY$) for vectors or matrices that depend on at least
one spatial variable, and serifed fonts (e.g. $\fb,\Db$\hsp) for those
that are entirely composed of ``spectral'' quantities. In the case of
dense, equal-area, whole-sphere sampling we have an 
approximation to eq.~(\ref{normalization}):
\be
\label{nohold}
\bY\bY\T\approx\Delta\Omega^{-1}\Ib\qquad \mbox{(case~1)}
,
\ee
where the elements of the $\Lpot\times\Lpot$-dimensional spectral
identity matrix~$\Ib$ are given by the Kronecker deltas
$\dllp\dmmp$. In the case of dense, equal-area, sampling over some closed
region~$R$, we find instead an approximation to the  $\Lpot\times
\Lpot$-dimensional ``spatiospectral localization matrix'':
\be
\label{nohold2}
\bY\bY\T\approx\Delta\Omega^{-1}\Db\qquad \mbox{(case 2)}
,
\ee
where the elements of~$\Db$ are those defined in
eq.~(\ref{Dlmlmpdef}). 

Let us now introduce the $\Lpot\times \Lpot$-dimensional matrix of
spectral Slepian eigenfunctions by 
\be
\Gb=\left(\barray{ccccc}
g_{00\hsp1}       & \cdots     & g_{00\hsp\alpha}  & \cdots &  g_{00\hsp\Lpot}\\
                 &            & \vdots           &        &       \\
\cdots           &            & g_{lm\hsp\alpha}    &        &\cdots \\
                 &            & \vdots           &        &       \\
g_{LL\hsp 1} & \cdots     & g_{LL\hsp\alpha}  & \cdots &g_{LL\hsp\Lpot}
\earray
\right)
.
\ee
This is the matrix that contains the eigenfunctions of the problem
defined in eq.~(\ref{fulleigen1}), which we rewrite as
\be
\Db\hsp\Gb=\Gb\hsp\blambda
,
\ee
where the diagonal matrix with the concentration eigenvalues is given by
\be
\blambda=\diag\left(\;\lambda_1\;\cdots\;\lambda_\alpha\;\cdots\;\lambda_{\Lpot}\;\right)
.
\ee
The spectral orthogonality relations of eq.~(\ref{orthogn}) are
\be
\label{sleportho}
\Gb\Trm\Gb=\Ib,\qquad
\Gb\Trm\Db\hsp\Gb=\blambda
,
\ee
where the elements of the $\Lpot\times\Lpot$-dimensional Slepian
identity matrix~$\Ib$ are given by the Kronecker deltas
$\dab$. We write the Slepian functions of eq.~(\ref{bandlg}) as
\be
\label{slepwrite}
\begg=\Gb\Trm\bY
\also
\bY=\Gb\hsp\begg
,
\ee
where the $\Lpot\times M$-dimensional matrix holding the sampled
spatial Slepian functions is given by
\be
\begg=\left(\barray{ccccc}
g_{1}(\rhat_1) & \cdots     & g_{1}(\rhat_j)  & \cdots &  g_{1}(\rhat_M)\\
                &            & \vdots           &        &       \\
\cdots          &            & g_\alpha(\rhat_j)    &        &\cdots \\
                &            & \vdots           &        &       \\
g_{\Lpot}(\rhat_1) & \cdots     & g_{\Lpot}(\rhat_j)  & \cdots &g_{\Lpot}(\rhat_M)       
\earray
\right)
.
\ee
Under a dense, equal-area, whole-sphere sampling, we will
recover the the spatial orthogonality of eq.~(\ref{orthog})
approximately as
\be
\label{ggt1}
\begg\hsp\begg\T\approx\Delta\Omega^{-1}\Ib\qquad \mbox{(case~1)}
,
\ee
whereas for dense, equal-area, sampling over a region~$R$ we will get,
instead, 
\be
\label{ggt2}
\begg\hsp\begg\T\approx\Delta\Omega^{-1}\blambda\qquad \mbox{(case 2)}
.
\ee
With this matrix notation we shall revisit both estimation problems of
the previous section.

\ssec{Problem~(i), revisited}

\sssec*{Spherical harmonic solution}

If we treat eq.~(\ref{synth}) as a noiseless inverse
problem in which the sampled data~$\beff$ are given but from which
the coefficients~$\fb$ are to be determined, we find that for dense,
equal-area, whole-sphere sampling, the solution 
\be
\label{inv0}
\fbh\approx\Delta\Omega\,\bY\hsp\beff
\qquad\mbox{(case~1)}
\ee
is simply the discrete approximation to the spherical harmonic
analysis formula~(\ref{expansion}). For dense, equal-area, regional
sampling we need to calculate
\be
\label{inv2}
\fbh\approx\Delta\Omega\,\Db^{-1}\hsp\bY\beff
\qquad\mbox{(case 2)}
.
\ee
Both of these cases are simply the relevant solutions to the familiar
overdetermined spherical harmonic inversion
problem\cite{Kaula67a,Menke89,Aster+2005a} for discretely sampled
data, i.e. the least-squares solution to eq.~(\ref{synth}),
\be
\label{inv1}
\fbh=(\bY\bY\T)^{-1}\bY\hsp\beff
,
\ee
for the particular cases described by
eqs~(\ref{nohold})--(\ref{nohold2}). In eq.~(\ref{inv2}) we
furthermore recognize the discrete version of eq.~(\ref{hatss}) with
$\eta=0$, the undamped solution to the minimum mean squared error
inverse problem posed in continuous form in
eq.~(\ref{variational}). From the continuous limiting 
case eq.~(\ref{hatss}) we thus discover the general form that
damping should take in regularizing the ill-conditioned inverse required in
eqs~(\ref{inv2})--(\ref{inv1}). Its principal property is that it
differs from the customary \textit{ad hoc} practice of adding small
values on the diagonal only.  Finally, in the most general and
admittedly most commonly encountered case of randomly scattered data
we require the Moore-Penrose pseudo-inverse   
\be
\label{invgen}
\fbh=\pinv(\bY\T)\hsp\beff
,
\ee
which is constructed by inverting the singular value decomposition
(svd) of~$\bY\T$ with its singular values truncated beyond where they
become vanishingly small\cite{Xu98}. Solving eq.~(\ref{invgen}) by
truncated svd is equivalent to inverting a truncated eigenvalue
expansion of the normal matrix~$\bY\bY\T$ as it appears in
eq.~(\ref{inv1}), as can be easily shown. 

\sssec*{Slepian basis solution}

If we collect the Slepian expansion coefficients of the
function~$f$ into the $\Lpot\times 1$-dimensional vector
\be
\label{tdef}
\tb=(\;f_{1}\;\cdots\; f_{\alpha}\; \cdots\; f_{\Lpot}\;)\Tit
,
\ee
the expansion~(\ref{bandlf1}) in the Slepian basis takes the form
\be
\label{synthslep}
\beff=\begg\T\tb=\bY\T\Gb\hsp\tb
,
\ee
where we used eqs~(\ref{sleportho})--(\ref{slepwrite}) to obtain the
second equality. Comparing eq.~(\ref{synthslep}) with
eq.~(\ref{synth}), we see that the Slepian expansion coefficients of a
function transform to and from the spherical harmonic coefficients as: 
\be\label{sleptoharm}
\fb=\Gb\hsp\tb
\also
\tb=\Gb\Trm\fb
.
\ee 
Under dense, equal-area, sampling with complete coverage
the coefficients in eq.~(\ref{synthslep}) can be estimated from
\be
\label{invg}
\tbh\approx\Delta\Omega\,\begg\hsp\beff
\qquad\mbox{(case~1)}
,
\ee
the  discrete, approximate, version of eq.~(\ref{bandlf2}). 
For dense, equal-area, sampling in a limited region~$R$ we get
\be
\label{invg2}
\tbh\approx\Delta\Omega\,
\blambda^{-1}\hsp\begg\hsp\beff
\qquad\mbox{(case 2)}
.
\ee
As expected, both of the solutions~(\ref{invg})--(\ref{invg2}) are again
special cases of the overdetermined least-squares solution
\be
\label{invg4}
\tbh=
(\begg\begg\T)^{-1}
\begg\hsp\beff
,
\ee
as applied to eqs~(\ref{ggt1})--(\ref{ggt2}).  We encountered
eq.~(\ref{invg2}) before in the continuous form of
eq.~(\ref{SGsolspec}); it solves the undamped minimum mean squared
error problem~(\ref{variational}).  The regularization of
this ill-conditioned inverse problem may be achieved by truncation of the
concentration eigenvalues, e.g. by restricting the size of the
$\Lpot\times\Lpot$-dimensional operator~$\begg\begg\T$ to its first
$N\DDD\times N\DDD$ subblock. Finally,  in the most general, 
scattered-data case, we would be using an eigenvalue-truncated version
of eq.~(\ref{invg4}), or, which is equivalent, form the
pseudo-inverse  
\be
\label{invg3}
\tbh=\pinv(\begg\T)\hsp\beff
.
\ee

The solutions~(\ref{inv0})--(\ref{inv1})
and~(\ref{invg})--(\ref{invg4}) are equivalent and differ only by the
orthonormal change of basis from the spherical harmonics to the
Slepian functions. Indeed, using eqs~(\ref{sleptoharm})
and~(\ref{slepwrite}) to transform eq.~(\ref{invg4}) into an equation
for the spherical harmonic coefficients, and comparing with
eq.~(\ref{inv1}) exposes the relation
\be\label{trivial}
\Gb(\begg\begg\T)^{-1}\Gb\Trm=(\bY\bY\T)^{-1}
,
\ee
which is a trivial identity for case~1 (insert eqs~\ref{nohold},
\ref{ggt1} and~\ref{sleportho}) and,  after substituting
eqs~(\ref{nohold2}) and~(\ref{ggt2}), entails
\be
\Gb \blambda^{-1}\Gb\Trm=\Db^{-1}
\ee
for case~2, which holds by virtue of
eq.~(\ref{sleportho}). Eq.~(\ref{trivial}) can also be
verified directly from eq.~(\ref{slepwrite}), which implies
\be
\bY\bY\T=
\Gb(\begg\begg\T)\hsp\Gb\Trm
.
\ee
The popular but labor-intensive procedure by which the unknown
spherical harmonic expansion coefficients of a scattered data set are
obtained by forming the Moore-Penrose pseudo-inverse as in
eq.~(\ref{invgen}) is thus equivalent to determining the truncated
Slepian solution of eq.~(\ref{invg2}) in the limit of continuous and
equal-area, but incomplete data coverage. In that limit, the generic
eigenvalue decomposition of the normal matrix becomes a specific
statement of the Slepian problem as we encountered it before, namely
\be
\bY\bY\T\hspm\Delta\Omega=\Ub\bsigma^2\Ub\Trm
\quad\rar\quad
\Db=\Gb\blambda\Gb\Trm
.
\ee
Such a connection has been previously pointed out for time
series\cite{Wingham92} and leads to the notion of ``generalized
prolate spheroidal functions''\cite{Bronez88} should the ``Slepian''
functions be computed from a formulation of the concentration problem
in the scattered data space directly, rather than being determined by
sampling those  obtained from solving the corresponding continuous
problem, as we have described here.  

Above, we showed how to stabilize the inverse problem of
eq.~(\ref{inv1}) by damping. We dealt with the case of
continuously available data only; the form in which it appears in
eq.~(\ref{hatss}) makes it clear that damping is hardly practical for
scattered data. Indeed, it requires knowledge of the complementary
localization operator~$\bar{\Db}$, in addition to being sensitive to
the choice of~$\eta$, whose optimal value depends implicitly on the
unknown signal-to-noise ratio\cite{Simons+2006b}. The data-driven
approach taken in eq.~(\ref{invgen}) is the more sensible
one\cite{Xu98}. We have  now seen that, in the limit of continuous
partial coverage, this corresponds to the optimal solution of the
problem formulated directly in the Slepian basis. It is consequently
advantageous to also work in the Slepian basis in case the  data
collected are scattered but closely collocated in some region of
interest. Prior knowledge of the geometry of this region and a prior
idea of the spherical harmonic bandwidth of the data to be inverted
allows us to construct a Slepian basis for the situation at hand, and
the problem of finding the Slepian expansion coefficients of the
unknown underlying function can be solved using
eqs~(\ref{invg4})--(\ref{invg3}). The measure within which this
approach agrees with the theoretical form of eq.~(\ref{invg2}) will
depend on how regularly the data are distributed within the region of
study, i.e on the error in the approximation~(\ref{ggt2}). But if
indeed the scattered-data Slepian normal matrix~$\begg\begg\T$ is
nearly diagonal in its first $N\DDD\times N\DDD$-dimensional block due
to the collocated observations having been favorably, if irregularly,
distributed, then eq.~(\ref{invg2}), which, strictly speaking,
requires no matrix inversion, can be applied directly. If this is not
the case, but the data are still collocated or we are only interested
in a local approximation to the unknown signal, we can
restrict~$\begg$ to its first~$N\DDD$ rows, prior to
diagonalizing~$\begg\begg\T$ or performing the svd of a
partial~$\begg\T$ as  necessary to calculate
eqs~(\ref{invg4})--(\ref{invg3}).  Compared to solving
eqs~(\ref{inv1})--(\ref{invgen}), the computational savings 
will still be substantial, as only when~$R\approx\Omega$ will the
operator~$\bY\bY\T$ be nearly diagonal. Truncation of the eigenvalues 
of~$\bY\bY\T$ is akin to truncating the matrix~$\begg\begg\T$ itself,
which is diagonal or will be nearly so. With the theoretically,
continuously determined, sampled Slepian functions as a
parametrization, the truncated expansion is easy to obtain and the
solution will be locally faithful within the region of
interest~$R$. In contrast, should we truncate~$\bY\bY\T$ itself,
without first diagonalizing it, we would be estimating a low-degree
approximation of the signal which would have poor resolution
everywhere.    


\sssec*{Bias and variance}

For completeness we briefly return to the expressions for the
mean squared estimation error of the damped spherical-harmonic and
the truncated Slepian function methods,
eqs.~(\ref{msefinal})--(\ref{SGmsefinal}), which we quoted for the
example of ``white'' signal and noise. Introducing the  
$\Lpot\times \Lpot$-dimensional spectral matrices  
\begin{subequations}
\be
\Hb=\blambda+\eta\hsp(\Ib-\blambda),
\ee
\be
\Vb=N\Hb^{-2} \blambda,\also
\Bb=\sqrt{S}\,\Hb^{-1}(\Ib-\blambda),
\ee
\end{subequations}
we handily rewrite the ``full'' version of eq.~(\ref{msefinal}) in two
spatial variables as the error covariance matrix
\be
\langle\epsilon(\rhat)\epsilon(\rhat')\rangle=
\begg\T\!\!\left(\Vb+\eta^2\Bb^2\right)\!\begg
.
\ee
We subdivide the matrix with Slepian functions into the
truncated set of the best-concentrated $\alpha=1\rar N\DDD$ and the
complementary set of remaining $\alpha=N\DDD+1\rar\Lpot$ functions,
as follows   
\be
\begg=
\big(\;\beggubar\T \;\; \beggbar\T\big)\T,
\ee
and similarly separate the eigenvalues, writing
\begin{subequations}
\ber
\blambdabar&=&\diag\left(\;\lambda_1\;\cdots\;\lambda_{N\DDD}\;\right),\\
\blambdaubar&=&\diag\left(\;\lambda_{N\DDD+1}\;\cdots\;\lambda_{\Lpot}\right)
.
\eer
\end{subequations}
Likewise, the identity matrix is split into two parts, $\Ibbar$ and
$\Ibubar$. If we now also redefine 
\begin{subequations}
\be
\Wbbar=N\blambdabar^{-1},\also
\Cbbar=\sqrt{S}\,\Ibbar,
\ee
\be
\Wbubar=N\blambdaubar^{-1},\also
\Cbubar=\sqrt{S}\,\Ibubar,
\ee
\end{subequations}
the equivalent version of eq.~(\ref{SGmsefinal}) is readily
transformed into the full spatial error covariance matrix
\be
\langle\epsilon(\rhat)\epsilon(\rhat')\rangle=
\beggubar\T\Wbubar\hsp\beggubar
+
\beggbar\T\Cbbar^2\hsp\beggbar
.
\ee
In selecting the Slepian basis we have thus successfully separated the
effect of the variance and the bias on the mean squared reconstruction
error of a noisily observed signal. If the region of observation is a
contiguous closed domain~$R\subset\Omega$ and the truncation takes
place at the Shannon number~$N\DDD$, we have thereby identified the variance
as due to \textit{noise} in the region where data are available, and
the bias to \textit{signal} neglected in the truncated expansion ---
which, in the proper Slepian basis, corresponds to the regions over
which no observations exist.

Finally, we shall also apply the notions of discretely acquired data
to the solutions of problem~(ii), below.

\ssec{Problem~(ii), revisited}

We need two more pieces of notation in order to rewrite the
expressions for the spectral estimates~(\ref{SestSP})
and~(\ref{SMTdef}) in the ``pixel-basis''. First we construct the
$M\times M$-dimensional symmetric spatial matrix collecting the
fixed-degree Legendre polynomials evaluated at the angular distances
between all pairs of observations points, 
\be
\bP_l=\tlofp
\left(\barray{ccccc}
P_l(\rhat_1\cdot\rhat_1) & \cdots     & P_l(\rhat_1\cdot\rhat_j)  & \cdots &  P_l(\rhat_1\cdot\rhat_M)\\
                &            & \vdots           &        &       \\
\cdots          &            & P_l(\rhat_i\cdot\rhat_j)    &        &\cdots \\
                &            & \vdots           &        &       \\
P_l(\rhat_M\cdot\rhat_1) & \cdots     & P_l(\rhat_M\cdot\rhat_j)  & \cdots &P_l(\rhat_M\cdot\rhat_M)       
\earray
\right)
.
\ee
The elements of~$\bP_l$ are thus
$\sum_{m=-l}^l\Ylm(\rhat_i)\Ylm(\rhat_j)$,  by the addition
theorem, eq.~(\ref{additionSH}). And finally, we
define~$\bG_l^{\alpha}$, the $M\times M$ symmetric matrix with
elements given by  
\be \label{Kpix}
\left(\bG_l^{\alpha}\right)_{ij}=
\left(\frac{2l+1}{4\pi}\right)g_{\alpha}(\br_i)P_l(\br_i\cdot\br_j)
g_{\alpha}(\br_{j}).
\ee

\sssec*{The spherical periodogram}

The expression equivalent to eq.~(\ref{SestSP}) is now written as
\be \label{SestSP2}
\hat{S}_l\SP=\left(\frac{4\pi}{A}\right)
\frac{(\Delta\Omega)^2}{2l+1}\,
\bd\T\bP_l\,\bd,
\ee 
whereby the column vector~$\bd$ contains the sampled data as in the
notation for eq.~(\ref{beff}). This lends itself easily to
computation, and the statistics of
eqs~(\ref{SexpecSP})--(\ref{moderSP}) hold, approximately, for
sufficiently densely sampled data.  

\sssec*{The spherical multitaper estimate}

Finally, the expression equivalent to eq.~(\ref{SMTdef}) becomes
\be \label{Salphadef2}
\hat{S}_l\MT=\sum_{\alpha=1}^{\Lpot}\lambda_\alpha
\left(\frac{4\pi}{N\DDD}\right)\frac{(\Delta\Omega)^2}{2l+1}\,
\bd\T\bG_l^{\alpha}\bd.
\ee
Both eqs~(\ref{SestSP2}) and~(\ref{Salphadef2}) are quadratic forms,
earning them the nickname ``quadratic spectral
estimators''\cite{Mullis+91}. The key difference with the
maximum-likelihood estimator popular in
cosmology\cite{Bond+98,Oh+99,Hinshaw+2003}, which can also be 
written as a quadratic form\cite{Tegmark97b}, is that neither~$\bP_l$
nor~$\bG_l^\alpha$ depend on the unknown spectrum itself. In contrast,
maximum-likelihood estimation is inherently non-linear, requiring
iteration to converge to the most probable estimate of the power
spectral density\cite{Dahlen+2008}. 

The estimate~(\ref{Salphadef2}) has the statistical properties that we
listed earlier as eqs~(\ref{SMTexpec})--(\ref{Gammafin}). These
continue to hold when the data pixelization is fine enough to have
integral expressions of the kind~(\ref{pixelsum}) be exact.  As
mentioned before, for completely irregularly and potentially
non-densely distributed discrete data on the sphere, ``generalized''
Slepian functions\cite{Bronez88} could be constructed specifically
for the purpose of their power spectral estimation, and used to build
the operator~(\ref{Kpix}). 

\section{Conclusions}

What is the information contained in a bandlimited set of scientific
observations made over an incomplete, e.g. temporally or spatially
limited sampling domain? How can this ``information'', e.g. an
estimate of the signal itself, or of its energy density, be
determined from noisy data, and how shall it be represented? These
seemingly age-old fundamental questions, which have implications
beyond the scientific\cite{Slepian76}, had been solved --- some say,
by conveniently ignoring them --- heuristically, by engineers, well
before receiving their first satisfactory answers given in the
theoretical treatment by Slepian, Landau and
Pollak\cite{Slepian+61,Landau+61,Landau+62}; first for ``continuous''
time series, later generalized to the multidimensional and discrete
cases\cite{Slepian64,Slepian78,Bronez88}.  By the ``Slepian
functions'' in the title of this contribution, we have lumped
together all functions that are ``spatiospectrally'' concentrated,
quadratically, in the original sense of Slepian. In one dimension,
these are the ``prolate spheroidal functions'' whose popularity is as
enduring as their utility. In two Cartesian dimensions, and on the
surface of the unit sphere, their time is yet to come,  and we have
made a case for it in this study. 

The answers to the questions posed above are as ever relevant for the
geosciences of today. There, we often face the additional
complications of irregularly shaped study domains, scattered
observations of noise-contaminated potential fields, perhaps collected
from an altitude above the source by airplanes or satellites, and an
acquisition and model-space geometry that is rarely if ever
non-spherical. Thus the Slepian functions are especially suited for
geoscientific applications and to study any type of geographical
information, in general.

Two problems that are of particular interest in the geosciences, but
also further afield, are how to form a statistically ``optimal''
estimate of the signal giving rise to the data, and how to estimate
the power spectral density of such signal. The first, an inverse
problem that is linear in the data, applies to forming mass flux
estimates from time-variable gravity, e.g. by the \textsc{grace}
mission, or to the characterization of the terrestrial magnetic field
by satellites such as \textsc{champ} or \textsc{swarm}. The second,
which is quadratic in the data,  is of interest in studying the
statistics of the Earth's or planetary topography, and especially for
the cross-spectral analysis of gravity  and topography, which can
yield important clues about the internal structure of the planets. The 
second problem is also of great interest in cosmology,
where missions such as \textsc{wmap} and \textsc{planck} are mapping
the cosmic microwave background radiation, which is best modeled
spectrally to constrain models of the evolution of our universe.  

Slepian functions, as we have shown by focusing on the case of
spherical geometry, provide the mathematical framework to solve such
problems. They are a convenient and easily obtained doubly-orthogonal
mathematical basis in which to express, and thus by which to recover,
signals that are geographically localized, or incompletely (and
noisily) observed. For this they are much better suited than the
traditional Fourier or spherical harmonic bases, and they are more
``geologically intuitive'' than wavelet bases in retaining a firm
geographic footprint and preserving the traditional notions of
frequency or spherical harmonic degree. They are furthermore extremely
performant as data tapers to regularize the inverse problem of power
spectral density determination from noisy and patchy observations,
which can then be solved satisfactorily without costly
iteration. Finally, by the interpretation of the Slepian
functions as their limiting cases, much can be learned about the
statistical nature of such inverse problems when the data provided are
themselves scattered within a specific areal region of study. 


\acknowledgments

For this work I am most indebted to Tony Dahlen (1942--2007), for
sharing my interest in this subject and being my friend, teacher and
mentor until his untimely death. I also thank Mark Wieczorek and
Volker Michel for many enlightening discussions on this and similar
topics over the last several years. Financial support for this work
was provided by the U.~S.~National Science Foundation under Grants
EAR-0105387 and EAR-0710860, and by the Universit\'e Paris
Diderot--Paris~7 and the Institut de Physique du Globe de Paris in
St.~Maur-des-Foss\'es, where this contribution was completed. Computer
algorithms are made available on \url{www.frederik.net}.  


\end{document}